\documentclass[osajn,pra,a4paper,reprint,showpacs,nofootinbib,superscriptaddress,floatfix,floats]{revtex4-1}
\usepackage{pifont}
\usepackage{amsmath}
\usepackage{amssymb}
\usepackage{amsfonts}
\usepackage{times,txfonts}
\usepackage{bm}
\usepackage{bbm}
\usepackage{dsfont}
\usepackage{bbold}
\usepackage{psfrag}
\usepackage{hyperref}
\hypersetup{
    colorlinks=true,
    citecolor=blue,
    linkcolor=blue,
    urlcolor=blue
}
\usepackage{color}
\usepackage{graphicx}
\usepackage{float}
\usepackage{epsfig}
\usepackage{verbatim}
\usepackage{epstopdf}
\usepackage{natbib}
\usepackage{simplewick}
\newcommand{\ket}[1]{\left\vert#1\right\rangle}

\newcommand{\ketbra}[2]{\left|{#1}\rangle \langle{#2}\right|}
\newcommand{\inner}[2]{\left\langle#1\kern-\nulldelimiterspace\left|#2\kern-\nulldelimiterspace\right.\right\rangle}

\newcommand{\half}{\frac{1}{2}}
\newcommand{\halfsmall}{\mbox{$\textstyle \half$}}

\begin{document}

\title{Effective Formalism for Open Quantum System Dynamics: Time-coarse-graining Approach}

\author{Chang-Woo Lee}
\email{changwoolee@kias.re.kr}
\affiliation{School of Computational Sciences, 85 Hoegi-ro, Dongdaemun-gu, Seoul 130-722, Korea}
\author{Changsuk Noh}
\email{changsuk@kias.re.kr}
\affiliation{School of Physics, 85 Hoegi-ro, Dongdaemun-gu, Seoul 130-722, Korea}
\author{Jaewan Kim}
\affiliation{School of Computational Sciences, 85 Hoegi-ro, Dongdaemun-gu, Seoul 130-722, Korea}

\date{\today}

\begin{abstract}
We formulate an effective-description framework for the dynamics of open quantum systems by extending the time-coarse-graining formalism to open systems. 
Our coarse-graining procedure efficiently removes high-frequency processes which are responsible for coherences between lower- and upper-manifold states and are irrelevant when considering only low-energy dynamics.
We investigate the regime of validity of the resulting coarse-grained master equation by applying it to multi-level atoms driven by far-detuned lasers. 
Except for such high-frequency coherences, 
we find good agreement between the exact and coarse-grained dynamics 
unless the driving lasers are too strong or the initial high-frequency coherences are sizable.
\end{abstract}

%\pacs{03.67.Mn, 42.50.Dv, 42.50.Xa, 42.50.-p}

\maketitle

\section{Introduction}
An \emph{effective description} of a physical system can help to gain insight into the structural and/or dynamical properties of the system. 
It can also help reduce the effort in extracting information about the system by decreasing its complexity or by transforming it into a more comprehensible form.
Sometimes, an effective description is the only plausible way to understand a system---e.g., in most many-body problems in condensed matter or statistical physics, while
in other cases, it may lead to a \emph{universal} identification of certain features of the system as in renormalization group procedures. 
On a more practical side, effective descriptions can also be of help in engineering quantum states
\cite{Poyatos96,Clark03,Parkins06,Hartmann07,Diehl08,Verstraete09,Kastoryano11,Torre13}.  

\emph{Adiabatic elimination} (AE) is a formalism often employed in studying quantum optical systems \cite{Gardiner04}.
If there are disparate---fast and slow---time scales in the dynamics of a system, 
AE provides an efficient procedure to adiabatically eliminate quantized levels that give rise to fast oscillations.
This method is, however, rather difficult to use and often requires tedious steps; accordingly, an easier and more practical version sharing the same spirit has recently been proposed \cite{Reiter12}.
Most AE methods, though, require definite knowledge of high- and low-energy manifolds, i.e., 
those to be removed and retained, respectively.
Therefore, AE approaches are often difficult to apply to a tangled multi-level system:
as its structure gets intricate---in particular, as 
two quantized levels hybridize---the distinction between excited and ground state manifolds becomes vague and the algebraic complexity grows rapidly at the same time.

There is a totally different approach to an effective description of the dynamics that adopts coarse-graining over fast time scales instead of bisecting the relevant Hilbert space and subsequently removing one of those \cite{Gamel10}.
Coarse-graining over fast time scales is equivalent to low-pass frequency filtering and consequently more efficiently separates the Hilbert space into its \emph{bona fide} high- and low-energy sections.
Differently from the AE methods, the time-coarse-graining (TCG) approach does not eliminate the excited states explicitly.
In this sense, TCG formalism is akin to the \emph{flow equation approach} \cite{Kehrein06}, which keeps the high-energy manifold 
while reducing effective couplings between the high- and low-energy manifolds.
In cases that one or more excited states need to be retained, 
e.g., when the initial state has a non-negligible 
portion of an excited state(s),
TCG holds obvious advantages over the AE methods that inevitably exclude those states.
If the contribution of some or all excited states turns out to be negligible, 
one can then safely remove such states after the coarse-graining.

In this work, we generalize the TCG formalism to open quantum systems and illustrate its performance and validity by applying it to simple 
prevalent examples of atomic systems driven by off-resonant laser fields. We find that a relatively simple and systematic formulation of the coarse-grained master equation [see Eq. \eqref{eq:meo7simple}] suffices to describe the `average' dynamics of such systems, given that one works in the perturbative regime with respect to the external driving fields, and the initial state does not contain a large amount of high-frequency coherences. This work is organized as follows. In Sect.~\ref{sec:tcgclosed} we review the TCG formalism for closed systems which is generalized to open systems in Sect.~\ref{sec:tcgopen}. It is then applied to systems typically found in quantum optics in Sect.~\ref{sec:examples}. First, we derive the time-coarse-grained master equation for a 4-level system driven by 4 off-resonant external fields which is the most general system that will be studied in this work. We then limit ourselves to first the simplest non-trivial case of a driven qubit and study the regime of validity of the effective master equation and extend these results to more complicated systems. 
We conclude by summarizing our findings and stating the simplified coarse-grained master equation that closely resembles the closed-system version found in \cite{Gamel10}.

\section{Time-coarse-graining Formalism for a Closed System}
\label{sec:tcgclosed}

Before putting forward our open-system version of the TCG formalism,
we briefly review the TCG formalism for a closed system \cite{Gamel10}.
Let us begin with the definition of time-coarse-graining---or \emph{time-averaging} in the nomenclature of \cite{Gamel10}. The time-coarse-grained version of an operator $O(t)$ is defined as
\begin{equation}
\overline{O} (t) \equiv \int_{-\infty}^{\infty} dt' f(t-t') O(t')
\end{equation}
for some (low-pass) filter function $f(t)$ such that 
$ \int_{-\infty}^{\infty} dt' f(t') = 1$.
Note that the TCG of the time derivative of an operator is equivalent to the time derivative of its TCGed operator, i.e.,
$\overline{\dot{O}}(t) = \dot{\overline{O}} (t)$ 
if $f(t)\rightarrow 0$ as $t \rightarrow \pm \infty$, which is a reasonable locality assumption.

An equation of motion for a closed quantum system with Hamiltonian $H$ can be written in the form of the von Neumann equation
\begin{equation}
\frac{d \rho}{dt} = -i [H(t), \rho],
\label{eq:mec1}
\end{equation}
where $\rho$ is the density matrix of the system, which contains the complete information about the system.
We now introduce a (real) bookkeeping parameter $\lambda$ such that $H(t) \rightarrow \lambda H(t)$ in the above equation to keep track of the order of Hamiltonian, which will be set to unity at the end.
Its formal solution is described by a unitary evolution such that
\begin{equation}
\rho (t) = U (t) \, \rho_0 U^\dag (t),
\label{eq:sol_c}
\end{equation}
where $\rho_0 \equiv \rho(0)$ and $U(t)$ is the time-evolution operator
\begin{eqnarray}
U(t) &\equiv& \mathbf{T} e^{-i \int_0^t H(t') dt'} \nonumber\\
&=&  \mathds{1} - i\lambda \int_{0}^{t} \!\! \, dt_1 H(t_1)  + (-i\lambda)^2 \int_{0}^{t} \!\! dt_1 \!\! \int_{0}^{t_1} \!\! dt_2 \, H(t_1) H(t_2)  \nonumber \\
& &  + \, \cdots  \nonumber \\
&\equiv&  \mathds{1} + \lambda U_1 (t) + \lambda^2 U_2 (t) + \, \cdots ,
\label{eq:u_c}
\end{eqnarray}
where $\mathbf{T}$ is the time-ordering operator, $\mathds{1}$ the identity operator, and by construction
$U_k(t) = O(H^k)$.
Since $U(t)$ is unitary,
\begin{eqnarray}
U^{-1} (t) &=& U^\dag (t) = \mathbf{\tilde{T}} e^{+i \int_0^t H(t') dt'}  \nonumber \\
&=&  \mathds{1} + i\lambda \int_{0}^{t} \!\! \, dt_1 H(t_1) + (i\lambda)^2 \int_{0}^{t} \!\! dt_1 \!\! \int_{0}^{t_1} \!\! dt_2 \, H(t_2) H(t_1)   \nonumber \\
& &  + \, \cdots  \nonumber \\
&=&  \mathds{1} + \lambda U_1^\dag (t) + \lambda^2 U_2^\dag (t) + \, \cdots
\label{eq:uinv_c}
\end{eqnarray}
where $\mathbf{\tilde{T}}$ is the \emph{anti}-time-ordering operator.
Note that rearranging the order of integration gives 
$U_2^\dag (t)= U_1^2 (t)- U_2 (t)$,
which can be obtained more easily by comparing at each order of $\lambda$ the both sides of
\begin{eqnarray}
\mathds{1} &=& U(t) U^\dag (t) \nonumber\\
&=& [ \mathds{1} + \lambda U_1 (t) + \lambda^2 U_2 (t) + \, \cdots] [\mathds{1} + \lambda U_1^\dag (t) + \lambda^2 U_2^\dag (t) + \, \cdots], \nonumber
\end{eqnarray}
namely,
\begin{equation}
0 = U_1 + U_1^\dag, \quad 0 = U_2 + U_1 U_1^\dag + U_2^\dag, \quad \cdots .
\label{eq:u_rel_c}
\end{equation}
Also notice the following relation
\begin{equation}
i \dot{U}_n(t) = H(t) U_{n-1} (t), \quad - i \dot{U}_n^\dag (t) = U_{n-1}^\dag (t) H(t) ,
\label{eq:u_eom_c}
\end{equation}
which are evident from the definitions of $U_n(t)$ and $U_n^\dag(t)$. 

Now we apply the time-coarse-graining action to \eqref{eq:sol_c} and use \eqref{eq:u_c} and \eqref{eq:uinv_c} to obtain
\begin{eqnarray}
\overline{\rho (t)} &=& \overline{U (t) \, \rho_0 U^\dag (t)}
= \sum_{k=0}^{\infty} \lambda^k \sum_{j=0}^{k} \overline{U_{k-j} \rho_0 U_j^\dag} \nonumber \\
&\equiv& \sum_{k=0}^{\infty} \lambda^k \mathcal{E}_k [\rho_0] \equiv \mathcal{E}[\rho_0]
\label{eq:sol_ta_c}
\end{eqnarray}
with $\mathcal{E}_0 = \mathds{1}$.
Next we define the inverse operator $\mathcal{F}$ such that 
\begin{equation}
\rho_0 = \mathcal{E}^{-1} [\overline{\rho}] \equiv \mathcal{F}[\overline{\rho}] \equiv \sum_{k} \lambda^k \mathcal{F}_k [\overline{\rho}]  .
\end{equation}
Using the identity relation $\mathcal{F}[\mathcal{E}[\rho]] = \rho$, i.e.,
\begin{equation}
\sum_{k=0}^{\infty} \lambda^k \sum_{j=0}^{k} \mathcal{F}_j[\mathcal{E}_{k-j}[\rho]] = \lambda^0 \rho,
\end{equation}
and by comparing the both sides at each order of $\lambda$, we get the following relations
\begin{equation}
\mathcal{F}_0 = \mathcal{E}_0 = \mathds{1}, \quad \mathcal{F}_1 = - \mathcal{E}_1, \quad \mathcal{F}_2 = \mathcal{E}_1^2 - \mathcal{E}_2, \quad\cdots .
\end{equation}
Differentiating \eqref{eq:sol_ta_c} leads to
\begin{eqnarray}
i \dot{\bar{\rho}} (t) &=& i \dot{\mathcal{E}}[\rho_0] = i \dot{\mathcal{E}}[\mathcal{F}[\bar{\rho} (t)]] 
= i \sum_{k=0}^{\infty} \lambda^k \sum_{j=0}^{k} \dot{\mathcal{E}}_j [\mathcal{F}_{k-j}[\bar{\rho} (t)]]  \nonumber\\
&\equiv& \sum_{k=0}^{\infty} \lambda^k \mathcal{L}_k [\bar{\rho} (t)],
\end{eqnarray}
where
\begin{eqnarray}
\mathcal{L}_0 [\bar{\rho}] &=& i\dot{\mathcal{E}}_0 [\mathcal{F}_0[\bar{\rho}]] = 0, \nonumber\\
\mathcal{L}_1 [\bar{\rho}] &=& i\dot{\mathcal{E}}_0 [\mathcal{F}_1[\bar{\rho}]] + i\dot{\mathcal{E}}_1 [\mathcal{F}_0[\bar{\rho}]] = [\bar{H}, \bar{\rho}], \nonumber\\
\mathcal{L}_2 [\bar{\rho}] &=& \contraction{}{H}{}{U} H U_1\bar{\rho} + \contraction{}{H}{\rho}{U} H \bar{\rho} U_1^\dag
- \contraction{\rho}{U}{_1^\dag}{H} \bar{\rho} U_1^\dag H - \contraction{}{U}{_1  \bar{\rho}}{H} U_1 \bar{\rho} H,
\end{eqnarray}
and 
\begin{equation}
\contraction{}{P}{}{Q} P Q \equiv \overline{PQ} - \overline{P}\,\overline{Q}, \quad
\contraction{}{P}{\bar{\rho}}{Q} P \bar{\rho} Q \equiv \overline{P \bar{\rho} Q} - \overline{P}\bar{\rho}\overline{Q} .
\end{equation}
We rewrite the master equation up to the second order as
\begin{equation}
i \dot{\rho} = [H_\text{eff}, \rho] + \left\{ \halfsmall ( A-A^\dag), \rho \right\} + 
\contraction{}{H}{\rho}{U} H \bar{\rho} U_1^\dag - \contraction{}{U}{_1  \bar{\rho}}{H} U_1 \bar{\rho} H,
\label{eq:mec2}
\end{equation}
where 
$\{P,Q\} \equiv PQ+QP$,
$A \equiv \contraction{}{H}{}{U} H U_1,~ A^\dag \equiv \contraction{}{U}{_1^\dag}{H} U_1^\dag H$,
and
\begin{equation}
H_\text{eff} = \overline{H} + \halfsmall (A+A^\dag).
\label{eq:h_eff_c}
\end{equation}
Except this effective Hamiltonian, all the other terms in \eqref{eq:mec2} represent decoherence processes.

For a specific time-coarse-graining process, Ref.~\cite{Gamel10} adopted the following rule
\begin{equation}
\overline{e^{\pm i \omega_n t}} = 0, \quad
\overline{e^{\pm i (\omega_m + \omega_n) t}} = 0, \quad
\overline{e^{\pm i (\omega_m - \omega_n) t}} = {e^{\pm i (\omega_m - \omega_n) t}} .
\label{eq:tav_c}
\end{equation}
Here one can notice that TCG formalism can be regarded as an extended RWA. 
At this point, there are two remarks in store.
First, we would get the same result if we start by differentiating 
\begin{equation}
\overline{U^\dag (t)  \, \rho (t) U (t)} = \rho_0 ,
\end{equation}
and obtaining the master equation order by order  using
$\bar{\rho} (t) = \bar{\rho}_0 (t) +\lambda \bar{\rho}_1 (t)+ \lambda^2 \bar{\rho}_2 (t) + \cdots $.
Second,
for a time-coarse-graining (or frequency filtering) process we can adopt a different procedure from \eqref{eq:tav_c} (e.g. one in Ref.~\cite{Wang15}); 
even in that case, all the previous formulas except \eqref{eq:tav_c} are still valid. 

Now let us apply the above procedure to the following class of Hamiltonians:
\begin{equation}
H = H_0 + \sum_n h_n e^{i \omega_n t} + h_n^\dag e^{-i \omega_n t} 
\label{eq:h_harmonic}
\end{equation}
where $H_0$ is independent of time. 
After some algebra we obtain (the details can be found in Ref. \cite{Gamel10} )
\begin{equation}
i\, \dot{\overline{\rho}} = [H_\text{eff}, \overline{\rho}] + 
\sum_{m, n} \frac{2}{\omega_{mn}^-} \left[ \mathcal{D}_{h_m(t), h_n^\dag(t)} \overline{\rho} - \mathcal{D}_{h_m^\dag(t), h_n(t)} \overline{\rho} \right] ,
\label{eq:mec3}
\end{equation}
where $\mathcal{D}_{A, B} \,\rho \equiv A \rho B - \halfsmall (BA \rho + \rho BA)$,
$\frac{1}{\omega_{mn}^\pm} \equiv \half \left( \frac{1}{\omega_m} \pm \frac{1}{\omega_n} \right)$,
$h_n(t) = h_n e^{i \omega_n t}$, and
\begin{eqnarray}
H_\text{eff} &\equiv& H_0 + \sum_{m, n} \frac{1}{\omega_{mn}^+} [h_m, h_n^\dag] e^{i (\omega_m-\omega_n) t} \nonumber \\
&=& H_0 + \sum_{m, n} \frac{1}{\omega_{mn}^+} \left(h_m (t) h_n^\dag (t) - h_m^\dag (t) h_n (t) \right) .
\end{eqnarray}

%%%%%%%%%%%%%%%%%%%%%%%%%%%%%%%%%%%%%%%%%%%%%%%%
\section{Time-coarse-graining Formalism for an Open System}
\label{sec:tcgopen}
%%%%%%%%%%%%%%%%%%%%%%%%%%%%%%%%%%%%%%%%%%%%%%%%

A typical form of master equation in an open system is written as 
(the role of $\lambda$ in the preivous section is evident so we will drop it from now on)
\begin{equation}
\frac{d \rho}{dt} 
= -i [H, \rho] + \sum_{i} \left( \mathcal{J}\!_{L_i} \rho - \mathcal{K}_{L_i}  \rho \right),
\label{eq:meo1}
\end{equation}
where $\mathcal{J}$ and $\mathcal{K}$ are superoperators defined as
\begin{equation}
\mathcal{J}\!_{L} \, \rho \equiv L \, \rho L^\dag, \quad \mathcal{K}_L \, \rho \equiv \half \left( L^\dag L \,  \rho + \rho L^\dag L \right).
\end{equation}
Denoting $\mathcal{K}_\text{tot} = \sum_i \mathcal{K}_{L_i}$, we can then write
\begin{eqnarray}
e^{\mathcal{K}_\text{tot} t} \rho &=& \left[ 1+ \mathcal{K}_\text{tot} t + \half (\mathcal{K}_\text{tot} t)^2 + \cdots \right] \rho \nonumber \\
&=& \rho + (K \rho + \rho K)t + \half (K^2  \rho + 2 K \rho K + \rho K^2) t^2 + \cdots \nonumber \\
&=& e^{K t} \rho \, e^{K t}
\label{eq:transform_k}
\end{eqnarray}
where $K \equiv \half \sum_{i} L_i^\dag L_i^{}$.
Applying this operation to \eqref{eq:meo1} we get
\begin{eqnarray}
e^{K t} \dot{\rho} e^{K t} &=& -i \left( H_K \rho_K - \rho_K H_K^\dag  \right)  +
\sum_{i}  \mathcal{J}\!_{L_{i,K}} \rho \nonumber \\
 && - e^{K t} (K \rho + \rho K)  e^{K t}
 \label{eq:meo2temp}
\end{eqnarray}
where 
\begin{equation}
\rho_K \equiv e^{K t} \rho e^{K t}, \quad H_K \equiv e^{K t} H e^{-K t}, \quad L_{i,K} \equiv e^{K t} L_i e^{-K t}.
\end{equation}
Noticing that 
$d\rho_K /dt = e^{K t} \dot{\rho} e^{K t} + e^{K t} ( K \rho + \rho K)  e^{K t}$,
Eq. \eqref{eq:meo2temp} reduces to
\begin{equation}
\frac{d \rho_K}{dt} = -i \left( H_K \rho_K - \rho_K H_K^\dag  \right)  +
\sum_{i}  \mathcal{J}\!_{L_{i,K}} \rho_K .
\label{eq:meo2}
\end{equation}

Next we make a key assumption needed to make further progress:
\begin{equation}
\left[K, ~ L_i \right]= - \gamma_{i,K}  L_i /2
\label{eq:assump}
\end{equation}
for some (relaxation) constant $\gamma_{i,K} $.
Note that this assumption is valid in many cases encountered in practice.
Then by using the Baker-Hausdorff theorem and \eqref{eq:assump}, we get
\begin{equation}
L_{i,K} = L_i + [K, L_i] t + \frac{1}{2!} [K,[K, L_i]] t^2 + \cdots 
= e^{-\gamma_{i,K} t /2} L_i .
\end{equation}
Now we define a \emph{pseudo}-time-evolution operator such that
\begin{equation}
U \equiv \mathbf{T} e^{-i \int_0^t H_K (t') dt'}.
\end{equation}
One can see that $U$ is not a unitary operator  since $H_K$ is not Hermitian
and that
\begin{equation*}
U^{-1} = \mathbf{\tilde{T}} e^{+i \int_0^t H_K dt'}, \
U^{\dag} = \mathbf{\tilde{T}} e^{+i \int_0^t H_K^\dag dt'}, \
(U^{-1})^\dag =  \mathbf{T} e^{-i \int_0^t H_K^\dag dt'}.
\end{equation*}

Next we expand $U,~U^{-1}$, and $U^\dag$ according to the order of $H_K$
as in \eqref{eq:u_c} and  \eqref{eq:uinv_c}, whose detailed forms are given in Appendix A.
Using these pseudo-evolution operators,
we can now transform the master equation into the \emph{pseudo}-rotating frame
where the density matrix is defined as $\rho_U \equiv U^{-1} \rho_K (U^{-1})^\dag$:
\begin{equation}
\frac{d \rho_U}{dt} = U^{-1} \dot{\rho}_K (U^{-1})^\dag +  U^{-1} i (H_K \rho_K - \rho_K H_K^\dag) (U^{-1})^\dag.
\end{equation}
Plugging Eq. \eqref{eq:meo2} into the above, we get
\begin{eqnarray}
\frac{d \rho_U}{dt} &=&  \sum_{i}  U^{-1} \mathcal{J}\!_{L_{i,K}} \rho_K (U^{-1})^\dag 
= \sum_{i} L_{i,U}^{} \rho_U L_{i,U}^\dag \nonumber\\
&=& \sum_{i} \mathcal{J}\!_{L_{i,U}} \rho_U = \mathcal{J}_\text{tot} \rho_U ,
\label{eq:meo3}
\end{eqnarray}
where $L_{i,U} \equiv  U^{-1} L_{i,K} U$ 
and $\mathcal{J}_\text{tot} \equiv \sum_{i} \mathcal{J}\!_{L_{i,U}} $.
With $\rho_0 \equiv \rho(0) = \rho_K(0) = \rho_U(0)$,
its formal solution is given by
\begin{eqnarray}
\rho_U (t) &=& e^{\int_{0}^{t} \mathcal{J}_\text{tot}  dt} \rho_0 \\
&=& \rho_0 + \sum_{i}  \int_{0}^{t} \!\! \, dt_1 L_{i,U}(t_1)\rho_0 L_{i,U}^\dag (t_1) 
+ \frac{1}{2!} \sum_{i,j}  \int_{0}^{t} \!\! dt_1 \!\! \int_{0}^{t_1} \!\! dt_2 \, \nonumber \\
& & \times \, L_{i,U}(t_1) L_{j,U}(t_2) \rho_0 L_{j,U}^\dag (t_2) L_{i,U}^\dag (t_1) + \cdots . \nonumber
\label{eq:rho_U_sol}
\end{eqnarray}
Since we are interested in the weak-dissipation regime, we will use the following approximation
\begin{equation}
\rho_U (t) \approx \left(1+ \int_{0}^{t} \mathcal{J}_\text{tot}  dt \right) \rho_0 
= \left(1+ \mathcal{J}_\text{int} \right) \rho_0 
\label{eq:rho_U_Jint}
\end{equation}
with $\mathcal{J}_\text{int} \equiv \int_{0}^{t} \mathcal{J}_\text{tot}  dt $.
Subsequently, let us turn back to $\rho_K$
\begin{equation}
\rho_K (t) = U \rho_U (t) U^\dag = U (1+ \mathcal{J}_\text{int} ) \rho_0 U^\dag ,
\end{equation}
and time-coarse-grain it
\begin{equation}
\overline{\rho}_K (t) = \overline{U (1+ \mathcal{J}_\text{int} ) \rho_0 U^\dag } \equiv \mathcal{E} [\rho_0] = \sum_k \mathcal{E}_k [\rho_0] ,
\label{eq:rho_K_sol}
\end{equation}
where $\mathcal{E}_k = O(H^k)$.
Since the subsequent procedure is almost similar to the closed system case,
let us just briefly sketch it and leave the detailed derivation to Appendix A.  
We first expand $L_{i,U}$ and $\mathcal{J}_\text{int}$ as
\begin{equation}
L_{i,U} \equiv L_{i,0} + L_{i,1} + L_{i,2} +  \cdots, \quad
\mathcal{J}_\text{int} \equiv \mathcal{J}_0 + \mathcal{J}_1 + \mathcal{J}_2 + \cdots ,
\end{equation}
and next 
$\mathcal{E}$ (along with its derivative) and its inverse $\mathcal{F}$ at each order. 
Equipped with these, we get the time-coarse-grained master equation for $\overline{\rho}_K (t)$
and, by inverting the transformation \eqref{eq:transform_k} (up to the 2nd order),
we finally obtain our main result, namely,
\begin{eqnarray}
i\, \dot{\overline{\rho}}(t) 
&=& [H_\text{eff},\overline{\rho}] + i \sum_{i} \left( \mathcal{J}\!_{L_i} \overline{\rho} - \mathcal{K}_{L_i}  \overline{\rho} \right) 
+ \left\{ \halfsmall ( A-A^\dag), \overline{\rho} \right\}  \nonumber\\
&& 
+ \contraction{}{H}{\overline{\rho}}{U} H \overline{\rho} \tilde{U}_1^\dag 
- \contraction{}{U}{_1 \rho}{H} \tilde{U}_1 \overline{\rho} H
+ \contraction{}{H}{(}{\mathcal{J}} H (\tilde{\mathcal{J}}_1 \overline{\rho}) 
- \contraction{(}{\mathcal{J}}{_1 \overline{\rho})}{H} (\tilde{\mathcal{J}}_1 \overline{\rho}) H,
\label{eq:meo6}
\end{eqnarray}
where $H_\text{eff} $ is the same as \eqref{eq:h_eff_c} with $A \equiv \contraction{}{H}{}{U} H \tilde{U}_{1},~A^\dag \equiv \contraction{}{U}{_1}{H} \tilde{U}_1^\dag H$ this time
and the operators with tilde are defined as
\begin{equation}
\tilde{U}_1 \equiv e^{-K t} U_1 \, e^{K t} ,  \quad
\tilde{\mathcal{J}}_1 \overline{\rho} \equiv e^{-K t} (\mathcal{J}_1\overline{\rho}_K) \, e^{-K t}.
\end{equation}
In deriving \eqref{eq:meo6}, we adopted the frequency filtering as before [see Eq.~\eqref{eq:tav_c}] and assumed that the decay processes are slow enough such that
\begin{equation}
\overline{O_K} = \overline{e^{K t} O \, e^{K t}} \approx e^{K t} \overline{O} \, e^{K t}.
\end{equation}
Further assuming that the Hamiltonian takes the form of \eqref{eq:h_harmonic} and 
\begin{equation}
e^{K t} h_i \, e^{-K t} = e^{-\kappa_i t } h_i
\quad \text{for some real constant } \kappa_i,
\end{equation}
the above master equation becomes
\begin{align}
\dot{\overline{\rho}} &= -i[H_\text{eff}, \overline{\rho}] + 
\sum_{n} \mathcal{D}_{L_n, L_n^\dag} \overline{\rho} \nonumber \\
&-i\sum_{m, n} \frac{2}{\tilde{\omega}_{mn}^-} \left[ \mathcal{D}_{h_m(t), h_n^\dag(t)} \overline{\rho} - \mathcal{D}_{h_m^\dag(t), h_n(t)} \overline{\rho} \right] \nonumber\\
& -i  \contraction{}{H}{(}{\mathcal{J}} H (\tilde{\mathcal{J}}_1 \overline{\rho}) 
+i \contraction{(}{\mathcal{J}}{_1 \overline{\rho})}{H} (\tilde{\mathcal{J}}_1 \overline{\rho}) H ,
\label{eq:meo7}
\end{align}
where $\tilde{\omega}_i \equiv \omega + i \kappa_i$,
$\frac{1}{\tilde{\omega}_{mn}^\pm} \equiv \half \left( \frac{1}{\tilde{\omega}_m} \pm \frac{1}{\tilde{\omega}_n^\ast} \right)$,
and
\begin{equation}
\label{eq:h_eff_o_1}
H_\text{eff} \equiv H_0 + \sum_{m, n} \frac{1}{\tilde{\omega}_{nm}^+} \left(h_m (t) h_n^\dag (t) - h_m^\dag (t) h_n (t) \right) .
\end{equation}
Additionally, the last two terms of \eqref{eq:meo6} and \eqref{eq:meo7} involve two-step (unitary plus nonunitary) processes  which lead to minute overall effect on the evolution. 
It turns out that for $H_0 = 0$ at least, these terms can be ignored with negligible loss in accuracy and
one can use (in most cases) the following simpler master equation
\begin{align}
\label{eq:meo7simple}
\dot{\overline{\rho}} &= -i[H_\text{eff}, \overline{\rho}] + 
\sum_{n} \mathcal{D}_{L_n, L_n^\dag} \overline{\rho} \nonumber \\
&-i\sum_{m, n} \frac{2}{\tilde{\omega}_{mn}^-} \left[ \mathcal{D}_{h_m(t), h_n^\dag(t)} \overline{\rho} - \mathcal{D}_{h_m^\dag(t), h_n(t)} \overline{\rho} \right].
\end{align}
Via numerical illustrations, the performances of and the comparison between \eqref{eq:meo7} and \eqref{eq:meo7simple} will be presented in the following section.

%%%%%%%%%%%%%%%%%%%%%%%%%%%%%%%%%%%%%%%%%
\section{Examples}
\label{sec:examples}
\begin{figure}[ht]
\includegraphics[width=0.65\columnwidth]{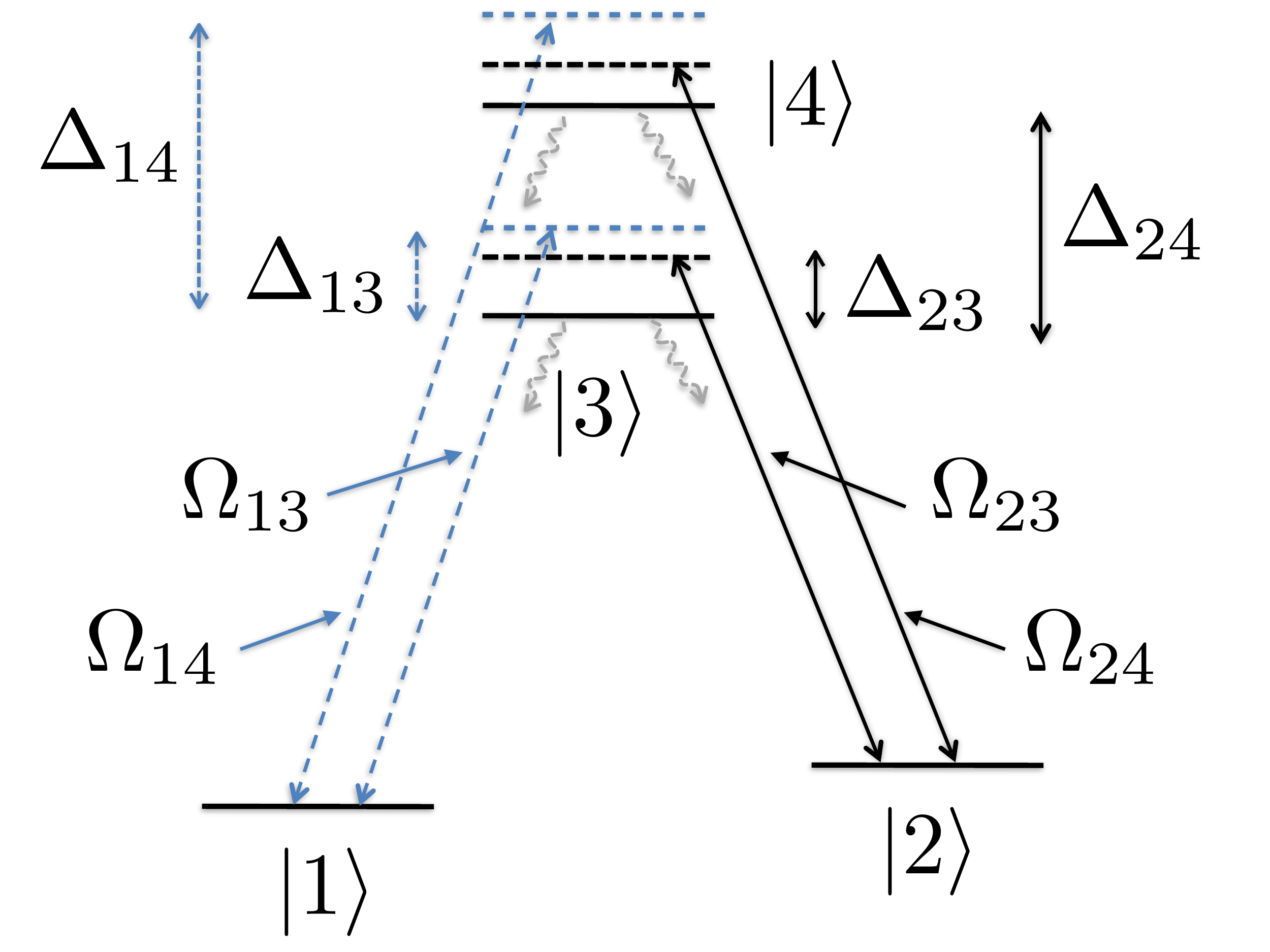}
\caption{Schematic illustration of a general setup to be studied in this section. A 4-level atom with double-lambda structure is driven by 4 far-detuned laser fields.
} 
\label{fig:4lvdiagram}
\end{figure}

%%%%%%%%%%%%%%%%%%%%%%%%%%%%%%%%%%%%%%%%%%
To illustrate the validity and shortcomings of the time-coarse-grained master equation, we will go through three examples of increasing complexity. We will start by stating the time-coarse-grained master equation for a four-level double-$\Lambda$ system depicted in Fig.~\ref{fig:4lvdiagram}, which will
encompass---by varying the parameters---all the examples illustrated in this section. The derivation is reproduced in Appendix B, which will make it clear that the master equation can be readily generalized to more complex systems. 
All the calculations were carried out using QuTiP \cite{Johansson12}.

%-----------------------------Figure 1---------------------------
%-----------------------------Figure 1---------------------------

In the interaction picture, the Hamiltonian for the four level system can be written as
\begin{align}
\label{Ham:4lv}
H &= \frac{\Omega_{13}}{2}|1\rangle\langle3|e^{i\Delta_{13}t} + \frac{\Omega_{14}}{2}|1\rangle\langle4|e^{i\Delta_{14}t} \nonumber \\ 
&+ \frac{\Omega_{23}}{2}|2\rangle\langle3|e^{i\Delta_{23}t} + \frac{\Omega_{24}}{2}|2\rangle\langle4|e^{i\Delta_{24}t} + \text{h.c.} 
\nonumber \\
 &\equiv \sum_{\langle i,j\rangle} h_{ij}(t) + h_{ij}^\dagger (t), 
\end{align}
where $\langle i,j\rangle \in \{ (1,3), (1,4), (2,3), (2,4)  \}$.
Going to the \emph{decaying} frame,
\begin{align}
H_K = \sum_{\langle i,j \rangle} h_{ij}e^{i\tilde{\Delta}_{ij}t} + h_{ij}^\dagger e^{-i\tilde{\Delta}_{ij}t},
\end{align}
where $h_{ij} = (\Omega_{ij}/2) |i\rangle\langle j|$ and $\tilde{\Delta}_{ij} = \Delta_{ij} + i\kappa_{ij}$; $\kappa_{13} = \kappa_{23} = \gamma_3$ and $\kappa_{14} = \kappa_{24} = \gamma_4$, where $\gamma_3 = (\gamma_{31}+\gamma_{32})/2$ and $\gamma_4 = (\gamma_{41}+\gamma_{42})/2$. Furthermore, $L_{1,K} = \gamma_{31}e^{-\gamma_3 t}|1\rangle\langle3|$, $L_{2,K} = \gamma_{32}e^{-\gamma_3 t}|2\rangle\langle3|$, $L_{3,K} = \gamma_{41}e^{-\gamma_4 t}|1\rangle\langle4|$, and $L_{4,K} = \gamma_{42}e^{-\gamma_4 t}|2\rangle\langle4|$. 
The first order pseudo-evolution operator $U_1 = -i\int_0^t dt' H_K (t')$ is easily calculated as
\begin{align}
U_1 = \sum_{\langle i,j\rangle} \frac{1}{\tilde{\Delta}_{ij}}\left( e^{-i\tilde{\Delta}_{ij}t}h_{ij}^\dagger -e^{i\tilde{\Delta}_{ij}t}h_{ij} \right) -  \frac{1}{\tilde{\Delta}_{ij}}\left( h_{ij} - h_{ij}^\dagger \right) .
\end{align}

With the assumption that the decaying terms $e^{-\gamma_i t}$ are unaffected by the time-coarse-graining procedure, %$U_1$ 
$\tilde{U}_1$ is obtained simply by the replacement $\tilde{\Delta}_{ij} \rightarrow \Delta_{ij}$ in the exponentiated factors only, i.e.,
\begin{align}
\tilde{U}_1 = \sum_{\langle i,j\rangle} \frac{1}{\tilde{\Delta}_{ij}}\left( e^{-i\Delta_{ij}t}h_{ij}^\dagger -e^{i\Delta_{ij}t}h_{ij} \right) -  \frac{1}{\tilde{\Delta}_{ij}}\left( h_{ij} - h_{ij}^\dagger \right) ,
\end{align}
and subsequently
\begin{align}
H\tilde{U}_1 =& \sum_{\langle i,j \rangle}\sum_{\langle k,l\rangle} \left[ h_{ij} e^{i\Delta_{ij}t} + h_{ij}^\dagger e^{-i\Delta_{ij}t} \right]\nonumber \\ &\times \left[ \frac{1}{\tilde{\Delta}_{kl}}\left( e^{-i\Delta_{kl}t}h_{kl}^\dagger -e^{i\Delta_{kl}t}h_{kl} \right) -  \frac{1}{\tilde{\Delta}_{kl}}\left( h_{kl} - h_{kl}^\dagger \right) \right].
\end{align}
Similarly as in the closed-system case, we coarse-grain out all $\exp(\pm i \Delta_{ij}t)$ terms along with their sum-frequency terms and keep only their difference-frequency ones.
We simply state the resulting master equation here and leave the derivation to Appendix B.
The resultant master equation can be written in the form
\begin{align}
\dot{\overline{\rho}} = -i\left[ H_{\rm eff}, \overline{\rho} \right] + \left( \mathcal{L}_{\rm deph} + \mathcal{L}_{\rm diss}+\mathcal{L}_{\rm jump}\right) \overline{\rho},
\end{align}
with
\begin{align}
H_{\rm eff} = \sum_{mn} \frac{1}{\tilde{\Delta}_{nm}^+} 
\left(h_m (t) h_n^\dag (t) - h_m^\dag (t) h_n (t) \right) ,
\label{eq:h_eff_o_2}
\end{align}
where the single indices $m$ and $n$ replace double indices $ij$ and $kl$. 
The `dephasing' term---it also includes `dissipative' terms, but we will call this \emph{dephasing} in analogy with that  of the closed-system equation---is
\begin{align}
\mathcal{L}_{\rm deph} = -i \sum_{m,n}\frac{2}{\tilde{\Delta}_{mn}^-} \left( \mathcal{D}_{h_m(t),h_n^\dag(t)} \overline{\rho}  - \mathcal{D}_{h_m^\dag(t),h_n(t)} \overline{\rho} \right),
\label{eq:L_deph}
\end{align}
with
\begin{align}
 \frac{1}{\tilde{\Delta}_{nm}^\pm} = \frac{1}{2}\left(\frac{1}{\tilde{\Delta}_{n}} \pm \frac{1}{\tilde{\Delta}_{m}^*}\right).
\end{align}
The `dissipation' terms are 
\begin{align}
\mathcal{L}_{\rm diss} = \sum_n\mathcal{D}_{L_n,L_n^\dagger} \overline{\rho}
\label{eq:L_diss}
\end{align}
and the `jump' terms  
\begin{align}
\mathcal{L}_{\rm jump}  = \sum_{n} \left[ L_n\bar{\rho} \left[ h_{\Delta,n}(t), L_n^\dagger \right],  h^\dagger (t) \right] + {\rm H.c.}
\label{eq:L_jump}
\end{align}
where $h(t) = \sum_m h_m(t)$ and
$h_{\Delta,n}(t) = \sum_m h_m(t)/[\tilde{\Delta}_m^* (\tilde{\Delta}_m^*+i\gamma_{n,K})]$.

%-----------------------------Figure 2---------------------------
\begin{figure}[ht]
	\includegraphics[width=1.0\columnwidth]{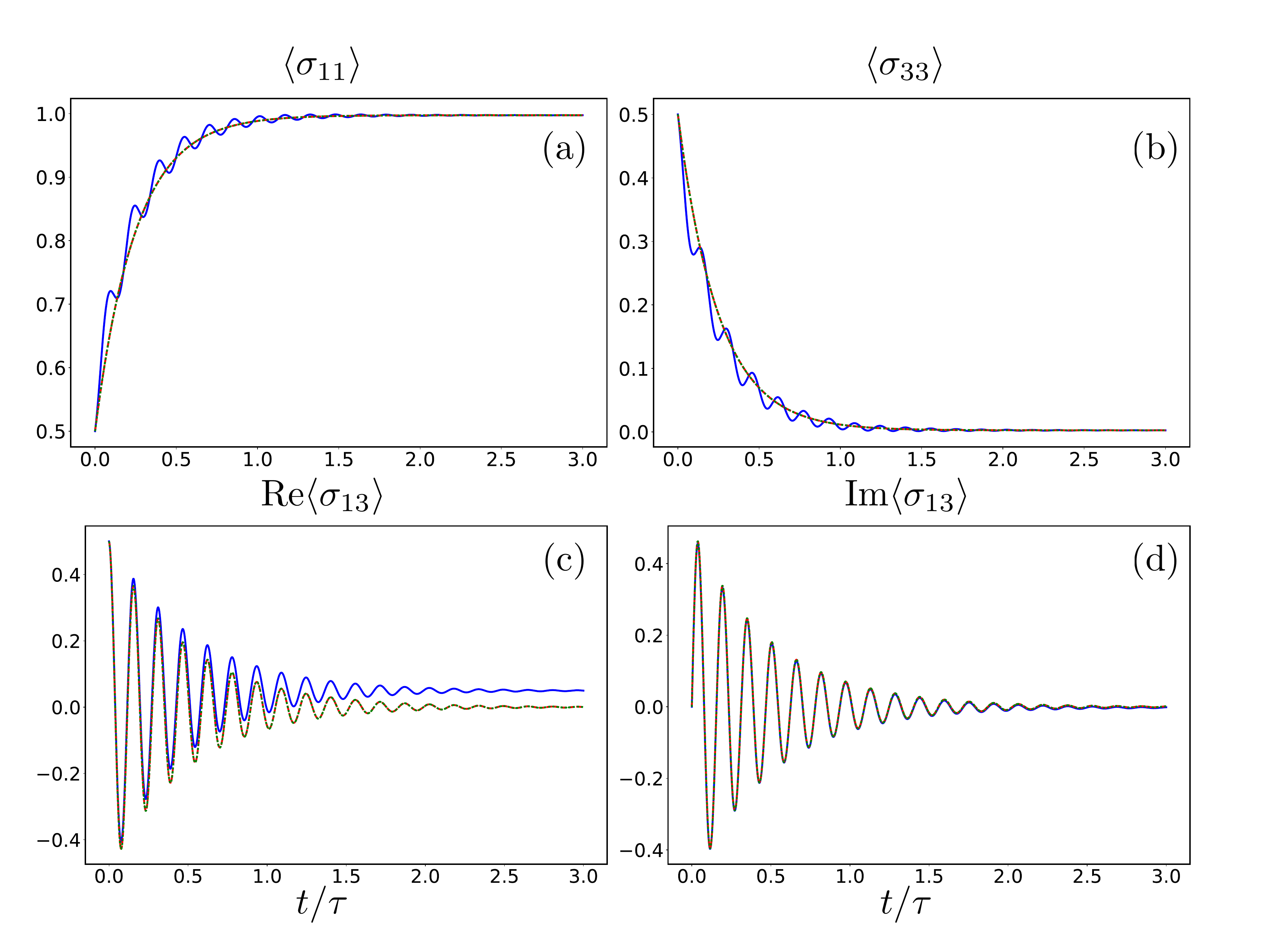}
	\caption{Time evolution of (a) $\langle\sigma_{11 }\rangle$, (b) $\langle\sigma_{33}\rangle$, (c) the real part of $\langle\sigma_{13}\rangle$, and (d) the imaginary part of $\langle\sigma_{13}\rangle$. Solid blue curves are the exact results, dashed red curves are the results obtained from the time-coarse grained master equation, and the dotted green curves are obtained by neglecting the `jump' terms in the coarse-grained dynamics. Here, $\Delta = 1$, $\Omega_{13} = 0.1$, $\gamma_{31} = 0.1$, and $\tau \equiv 4|\tilde{\Delta}_{13}|^2/\Delta_{13}|\Omega_{13}|^2$. }
	\label{fig:2lv}
\end{figure}
%-----------------------------Figure 2---------------------------

\subsection{Two-level System}
\label{sec:2level} 
Let us first consider the simplest nontrivial case, where only the states $|1\rangle$ and $|3\rangle$ are involved. 
All the parameters in Fig.~\ref{fig:4lvdiagram} are zero except those containing both 1 and 3 in  the subscript and $\Omega_{13}/\Delta_{13} \ll 1$ is assumed.  By defining 
$h_{13}(t) \equiv \Omega_{13}\exp(i\Delta_{13}t)\sigma_-/2$ and going to the rotating frame to remove the explicit time dependence, the master equation can be written explicitly as
\begin{align}
\dot{\overline{\rho}} =&-i\Delta_{13} \left[ |3\rangle\langle3|,\overline{\rho} \right] + \gamma_{31}\mathcal{D}_{\sigma_-,\sigma_+}\overline{\rho} \nonumber \\
& -i  \frac{\Delta_{13}|\Omega_{13}|^2}{4|\tilde{\Delta}_{13}|^2}\left[ [\sigma_-,\sigma_+],\overline{\rho}\right] \nonumber \\
&-\frac{\gamma_{31}|\Omega_{13}|^2}{4|\tilde{\Delta}_{13}|^2}  \left ( \mathcal{D}_{\sigma_-,\sigma_+}\overline{\rho} -   \mathcal{D}_{\sigma_+,\sigma_-}\overline{\rho}\right )\nonumber \\
&- \frac{\gamma_{31}|\Omega_{13}|^2}{4|\tilde{\Delta}_{13}|^2}\left[ \sigma_+,\sigma_-\overline{\rho}\left[ \sigma_-,\sigma_+\right] \right]_+ \nonumber \\
&+\frac{\gamma_{31}|\Omega_{13}|^2}{4|\tilde{\Delta}_{13}|^2}\left[ \sigma_-,\left[ \sigma_-,\sigma_+\right] \overline{\rho} \sigma_+\right]_+,
\label{eqn:twolv_eff}
\end{align}
where the last two terms correspond to the `jump' terms that contain $\mathcal{J}_1$ [see \eqref{eq:meo6}]. 

Now let us illustrate the performance of our TCG formalism 
by inspecting the time evolutions of various observables.
We will use the notation $\sigma_{ij} = |i\rangle \langle j|$ to name various observables. % through the plots of the time evolutions of density matrix entries. 
Figure \ref{fig:2lv} compares the exact dynamics (solid blue curves) with the time-coarse-grained dynamics (dashed red curves and dotted green curves) for $\Omega_{13}/\Delta_{13} = 0.1$, $\gamma_{31}/\Delta_{13} = 0.1$, and the initial state $|\psi_0\rangle \propto |1\rangle + |3\rangle$. 
Dashed red curves are obtained by solving the master equation \eqref{eqn:twolv_eff} while the dotted green curves are obtained by 
dropping the `jump' terms from it. We observe that the exact dynamics are well approximated by the coarse-grained equation and 
additionally that the `jump' terms contribute negligible corrections. This makes sense, 
since we are considering in the first place the regime of low dissipation rate and low excitation whereby the `jump' term in the original master equation plays little role. Also 
notice that the real value of the coherence term is not so well approximated by the coarse-grained dynamics [Fig.~\ref{fig:2lv}(c)]. We observe the same behaviour in more complicated examples below: the coherences involving the excited state manifold is in general not accurately approximated. This stems from the very nature of coarse-graining: such processes  
take place in the higher energy sector---or in other words oscillate at high frequencies---and hence are coarse-grained out in TCG formalism. In the coarse-grained dynamics, average values of such coherences always approach zero in the steady state.

\begin{figure}[t]
\includegraphics[width=1\columnwidth]{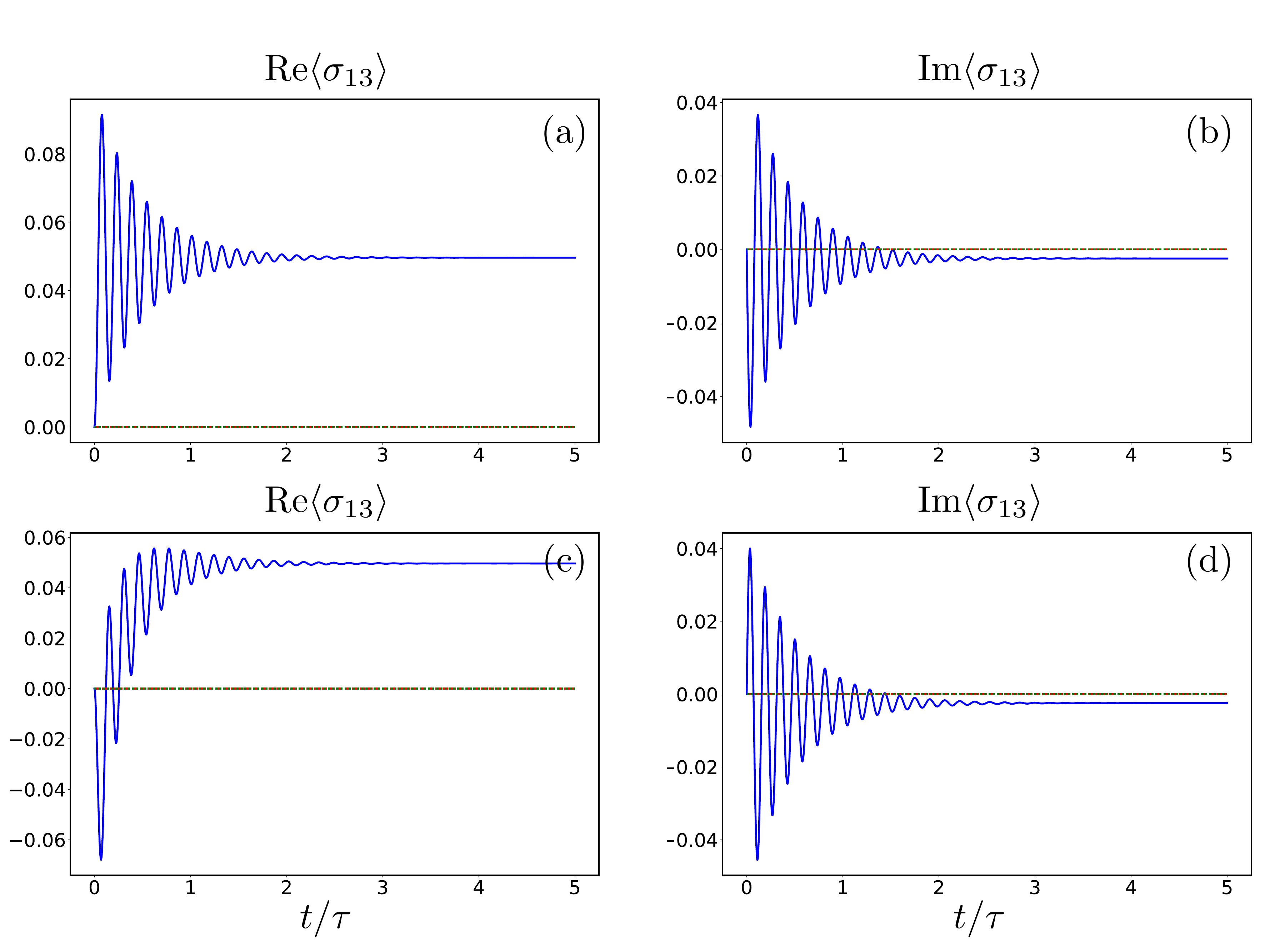}
\caption{Time evolution of the coherence $\langle \sigma_{13} \rangle$. Top row: the real (a) and imaginary (b) parts of $\langle\sigma_{13}\rangle$ for $|\psi_0 \rangle = |1\rangle$. Bottom row:   the real (c) and imaginary (d) parts of $\langle\sigma_{13}\rangle$ for $|\psi_0 \rangle= |3\rangle$. The same parameters are used as in Fig. \ref{fig:2lv}.}
\label{fig:2lv_init}
\end{figure}

For other initial states, the coherences fare even worse, while the populations remain well approximated. The extreme cases are when the initial coherence is zero as shown in Fig.~\ref{fig:2lv_init}. The top row is for $|\psi_0 \rangle = |1\rangle$ and the bottom row is for $|\psi_0 \rangle = |3\rangle$. We observe that when the initial coherence is zero it remains so and does not follow the exact evolution at all. 
 %-----------------------------Figure 3---------------------------
%-----------------------------Figure 3---------------------------

\begin{figure}[t]
\includegraphics[width=1\columnwidth]{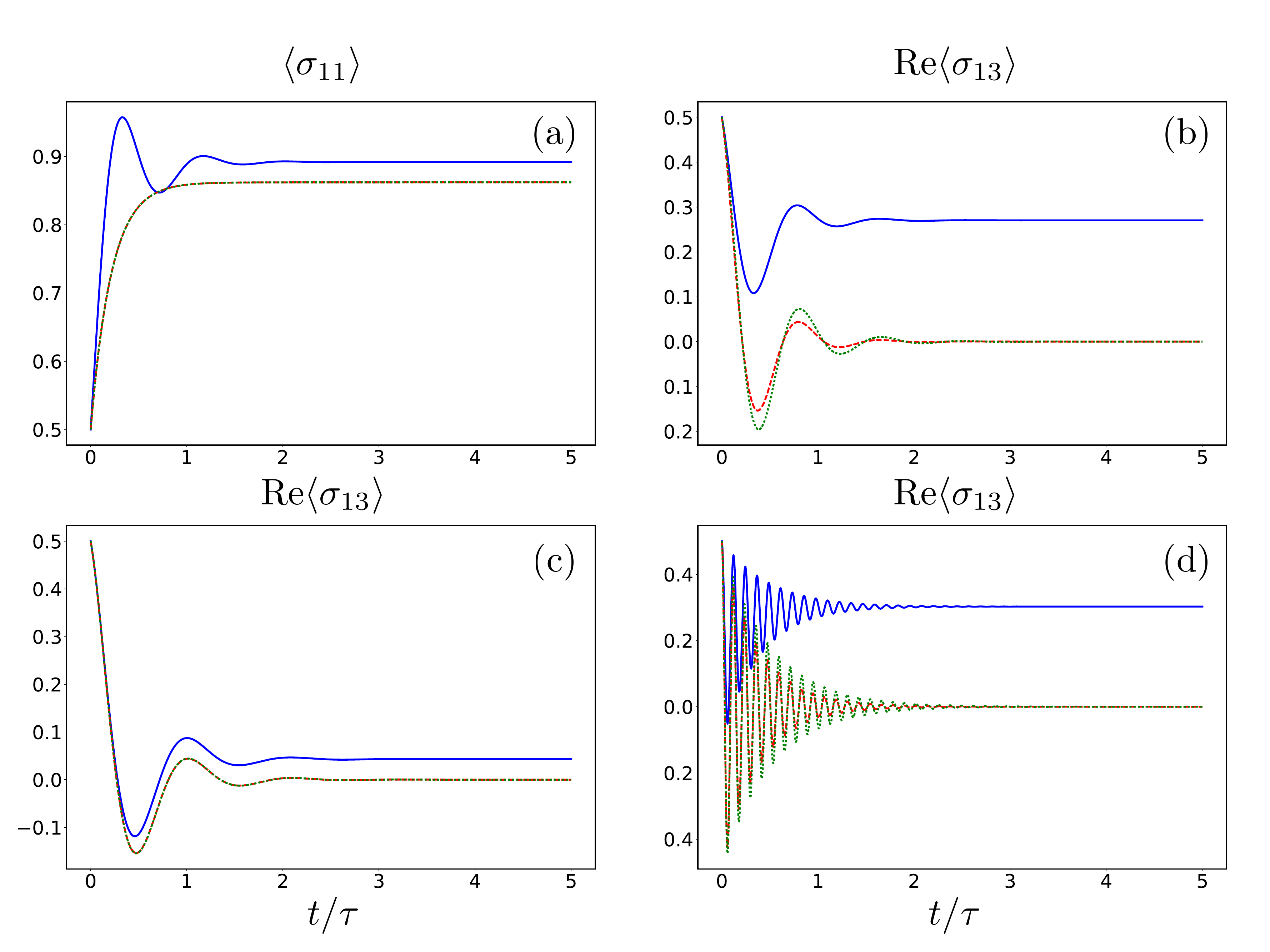}
\caption{Effects of the jump terms. (a) The ground state population for $\Omega_{13} = \gamma_{31} = 0.8$. The real part of the coherence is shown: (b) for the same set of parameters, (c) when $\Omega_{13} = 0.1$, and (d) when $\gamma_{31} = 0.1$. The parameters are in units of $\Delta_{13}$ and $|\psi_0\rangle \propto |1\rangle + |3\rangle$. }
\label{fig:2lv_jump}
\end{figure}

Next, we investigate the effects of the jump terms by comparing the dynamics with and without 
them in the coarse-grained master equation. 
Even in case of large perturbation parameters, i.e., $\Omega_{13}/\Delta_{13} = 0.8$ and $\gamma_{31}/\Delta_{13} = 0.8$, 
where the coarse-grained dynamics neither follows the true one nor predict the accurate steady-state value,
the jump terms produce no noticeable differences in the populations; see Fig.~\ref{fig:2lv_jump}(a) for the ground state population. 
The situation differs a bit as far as the coherence is concerned, as shown in Fig.~\ref{fig:2lv_jump}(b).
One might be tempted to deduce that 
differences of similar magnitude would be observed when the value of the coupling constant is reduced such that only the dissipation parameter is non-perturbative, but this is not the case as shown in Fig.~\ref{fig:2lv_jump}(c), where $\Omega_{13}/\Delta_{13}$ has been changed to 0.1. Similarly, reducing only the $\gamma_{31}/\Delta_{13}$ to 0.1 has no effect as shown in Fig.~\ref{fig:2lv_jump}(d). Note that we have only shown the real part of the coherence but the same level of discrepancy has been observed in the imaginary part as well.
These results tell us that we may simply ignore the jump terms in the coarse-grained master equation, since the effects of such terms only matter in the regime where TCG does not produce reliable results. The same conclusion is drawn from more complex systems studied below.
%-----------------------------Figure 4---------------------------
%-----------------------------Figure 4---------------------------

\subsection{Three-level Raman System}
\label{sec:raman}

%-----------------------------Figure 5---------------------------
\begin{figure}[t]
	\includegraphics[width=1\columnwidth]{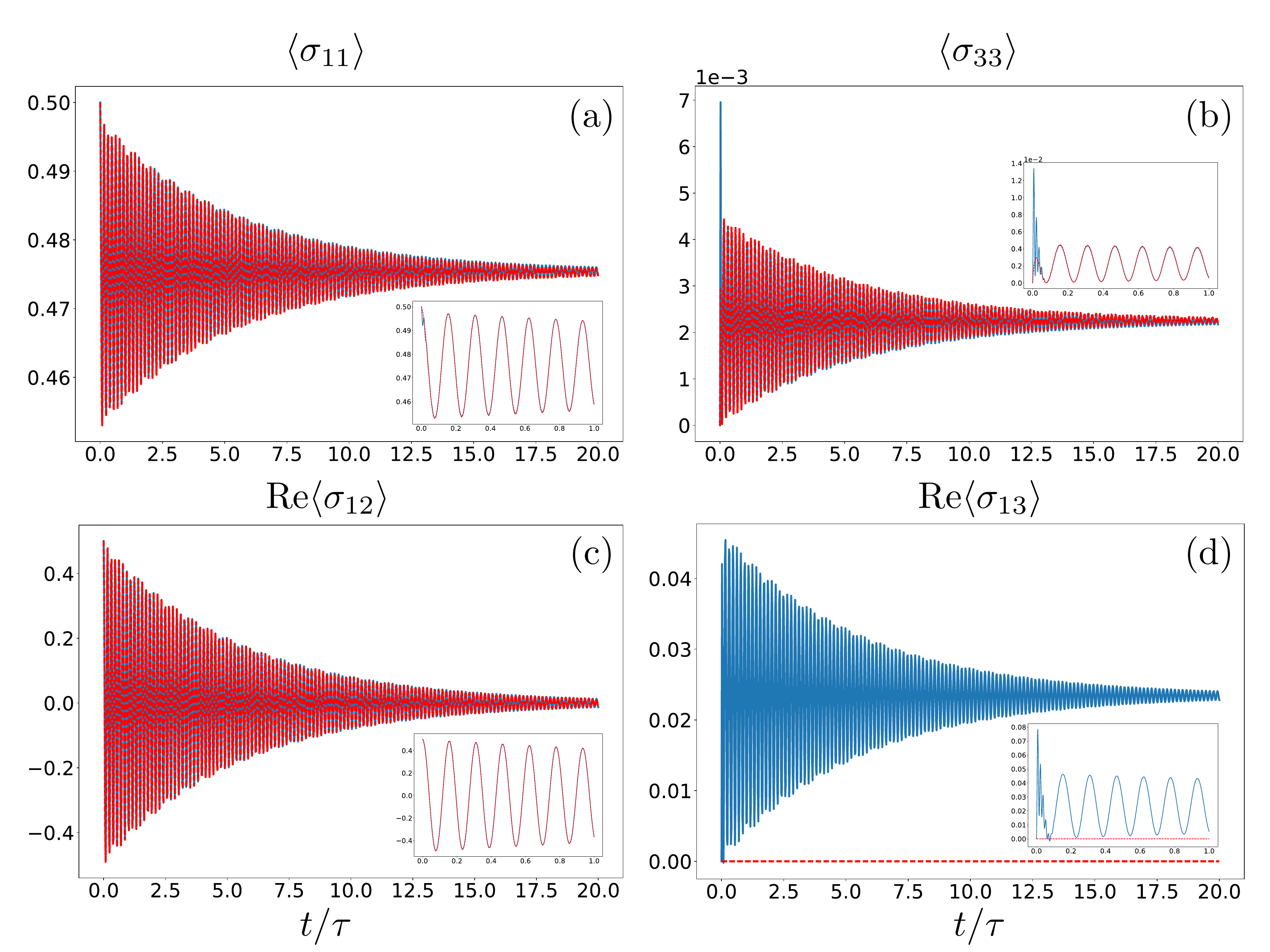}
	\caption{Time evolution for the three-level setup. The initial state is  $|\psi_0\rangle \propto |1\rangle + |2\rangle$ and parameters are $\Delta_1 = \Delta_2 = 1$, $\Omega_1 = 0.1$,  $\Omega_2 = 0.1$, and $\gamma_1 = \gamma_2 = 0.1$. Plots of (a) $\langle \sigma_{11} \rangle$, (b) $\langle \sigma_{33} \rangle$,  and the real parts of (c) $\langle \sigma_{12} \rangle$,  and (d) $\langle \sigma_{13} \rangle$ as functions of $t/\tau$, where $\tau = 4|\tilde{\Omega}_1|^2/\Omega_1^2 \Delta_1$. Solid blue curves represent exact dynamics whereas the dashed red curves represent coarse-grained dynamics. Insets are close-ups from $t=0$ to $\tau$.} 
	\label{fig:3lv}
\end{figure}
%-----------------------------Figure 5---------------------------

Next we 
proceed to the 3-level setup with 
$\ket{1}$ and $\ket{2}$ in the ground state manifold and only one excited state $\ket{3}$. This implies 
 that the low-energy scale coherence (in this case $\sigma_{12}$) is allowed. 
 The Hamiltonian for this system reads
\begin{equation}
H = \frac{\Omega_1}{2} \ketbra{1}{3} e^{i \Delta_1 t} + \frac{\Omega_2}{2} \ketbra{2}{3} e^{i \Delta_2 t} + \text{H.c.} .
\end{equation}
The operators describing decay from $\ket{3}$ to $\ket{i}$ ($i=1,~2$) are $L_i \equiv \sqrt{\gamma_i}\ketbra{i}{3}$. Then, $[(L_1^\dag L_1 + L_2^\dag L_2)/2, L_i] = (\gamma/2) L_i$ with $\gamma \equiv \gamma_1 + \gamma_2$,
and hence $L_{i,K} = L_{i,0} = e^{-\gamma t/2} L_i$ and 
\begin{equation}
H_K = h_1 e^{i \tilde{\Delta}_1 t} + h_2 e^{i \tilde{\Delta}_2 t}+ \text{H.c.} ,
\end{equation}
where $h_i \equiv ({\Omega_i}/{2}) \ketbra{i}{3}$ and $\tilde{\Delta}_i \equiv \Delta_i +i \gamma/2$.

Equation \eqref{eq:h_eff_o_2} leads to the effective Hamiltonian
\begin{eqnarray}
H_\text{eff} &=&  \sum_i  \frac{\Omega_i^2 \Delta_i}{4 \Delta_i^2 + \gamma^2} \left( \ketbra{i}{i} - \ketbra{3}{3} \right) \nonumber \\
&+& \left[ \frac{(\Delta_1 + \Delta_2) \Omega_1 \Omega_2}{8 ({\Delta}_1-i \gamma/2) ({\Delta}_2+i \gamma/2) }\ketbra{1}{2} e^{i \Delta_{12} t} + \text{H.c.} \right],
\end{eqnarray}
and Eq. \eqref{eq:L_deph} to the dephasing term 
\begin{align}
\mathcal{L}_{\rm deph} \cdot = -\frac{\Omega_1^2}{4}\frac{\gamma}{|\tilde{\Delta}_1|^2}\left[ \mathcal{D}_{|1\rangle\langle 3|,|3\rangle\langle 1|} \cdot  -\mathcal{D}_{|3\rangle\langle 1|,|1\rangle\langle 3|} \cdot \right]  \nonumber \\
-\frac{\Omega_2^2}{4}\frac{\gamma}{|\tilde{\Delta}_2|^2}\left[ \mathcal{D}_{|2\rangle\langle 3|,|3\rangle\langle 2|} \cdot  -\mathcal{D}_{|3\rangle\langle 2|,|2\rangle\langle 3|} \cdot \right]  \nonumber \\
-i\frac{\Omega_1 \Omega_2}{4}\frac{\tilde{\Delta}_2^* - \tilde{\Delta}_1}{\tilde{\Delta}_1\tilde{\Delta}_2^*}\left[ e^{i\Delta_{12}t}\mathcal{D}_{|1\rangle\langle 3|,|3\rangle\langle 2|} \cdot  \right. \nonumber \\ \left. -e^{-i\Delta_{12}t}\mathcal{D}_{|3\rangle\langle 1|,|2\rangle\langle 3|} \cdot \right] + {\rm H.c.}.
\end{align}
Finally, the dissipation terms are
\begin{align}
\mathcal{L}_{\rm diss} \cdot = \sum_n \mathcal{D}_{L_n,L_n^\dagger} \cdot
\end{align}
and
\begin{align}
\mathcal{L}_{\rm jump} \cdot = \sum_n \gamma_n \left[ |n\rangle\langle 3| \cdot [h_\Delta(t),|3\rangle\langle n|], h^\dagger(t) \right ] + {\rm H.c.},
\end{align}
where
\begin{align}
h(t) = \frac{\Omega_1}{2} e^{i\Delta_1t} |1\rangle\langle3|+\frac{\Omega_2}{2} e^{i\Delta_2t} |2\rangle\langle3|,
\end{align}
and
\begin{align}
h_{\Delta}(t) = \frac{\Omega_1}{2|\tilde{\Delta}_1|^2} e^{i\Delta_1t}|1\rangle\langle 3| + \frac{\Omega_2}{2|\tilde{\Delta}_2|^2} e^{i\Delta_2t}|2\rangle\langle 3| .
\end{align}

Note that by going to a rotating frame such that $|1\rangle\langle 3| \rightarrow |1\rangle\langle 3| e^{-i\Delta_1t}$ and $|2\rangle\langle 3| \rightarrow |2\rangle\langle 3| e^{-i\Delta_2t}$, both the exact and coarse-grained dynamics become time-independent. All the results in this subsection are computed in this rotating frame. Furthermore, as in the 2-level case the jump terms make negligible contribution and can be ignored entirely for the parameter regimes where the coarse-grained dynamics provide accurate approximations.

Figure \ref{fig:3lv} depicts the time-coarse-grained dynamics (dashed red curves) of various observables and compares them against the exact results (solid blue curves) for the initial state $\psi_0 \propto |1\rangle + |2\rangle$. As in the 2-level system, the populations are almost exactly matched for the parameters chosen (including the population in $|2\rangle$ which is not shown), except in the early stages (especially for $\sigma_{33}$). The ground-state coherence $\sigma_{12}$ is very well approximated, whereas the coherences involving the upper level are not, as expected from the study in the previous subsection. We find that the jump terms produce negligible difference throughout all the plots in this subsection (not shown).

\begin{figure}[t]
\includegraphics[width=1\columnwidth]{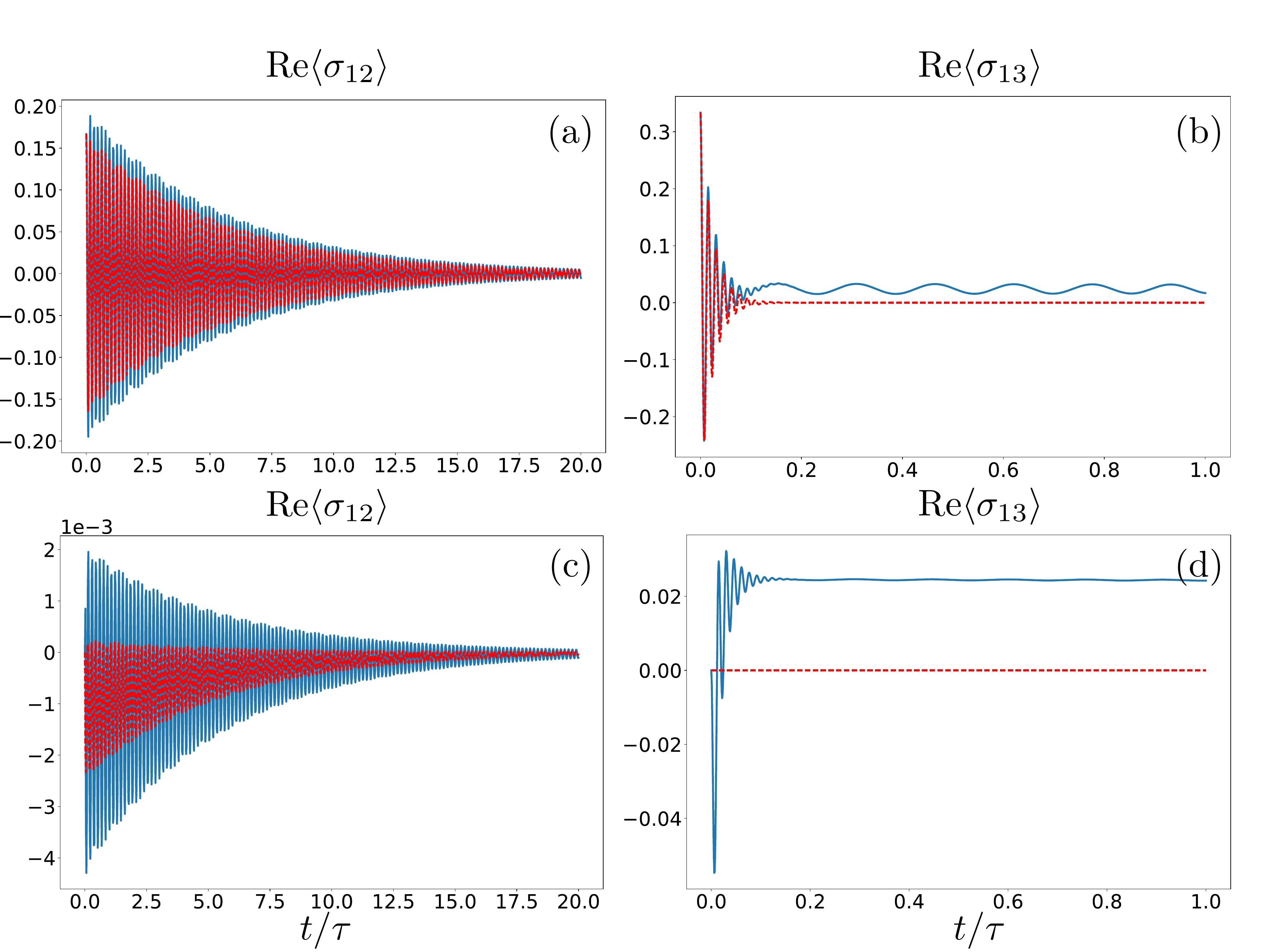}
\caption{Initial state dependence of the (real parts of) coherences.  The parameters are the same as in Fig. \ref{fig:3lv}. Top row: $|\psi_0\rangle \propto |1\rangle + |2\rangle + 2|3\rangle$. Bottom row: $\psi_0\rangle = |3\rangle$.}
\label{fig:3lv_init}
\end{figure}

In the above example, the initial state lies in the ground state manifold. 
What if the initial state contains some portion of excited states? We find that increasing the occupation of the excited state manifold has little effect on the populations, but significantly modifies the coherences as shown in Fig.~\ref{fig:3lv_init}. The top (bottom) row illustrates the results for $|\psi_0\rangle \propto |1\rangle + |2\rangle + 2|3\rangle$ ($|\psi_0\rangle = |3\rangle$). The ground-state coherence is less accurately approximated as the upper level population in the initial state is increased, whereas the high-frequency coherences, $\sigma_{13}$ and $\sigma_{23}$ (not shown), are better approximated if there is a non-zero initial coherence between the ground states and the excited state---in accordance with the previous subsection. The other coherences not shown in the figure exhibit similar behaviour.
%-----------------------------Figure 6--------------------------
%-----------------------------Figure 6---------------------------

\subsection{Four-level system}
The full master equation for the four-level system is straightforward to obtain but cumbersome to write down. Therefore, we will only show results obtained from the master equation here. One noticeable difference to the previous examples is that the 4-level setup is 
intrinsically time-dependent in the sense that the Hamiltonian cannot be written in a time-independent form by going to any rotating frame. We work in a rotating frame such that all terms except the $|2\rangle\langle 4|$ term and its Hermitian conjugate becomes time independent in Eq.~(\ref{Ham:4lv}).

\begin{figure}[t]
\includegraphics[width=1\columnwidth]{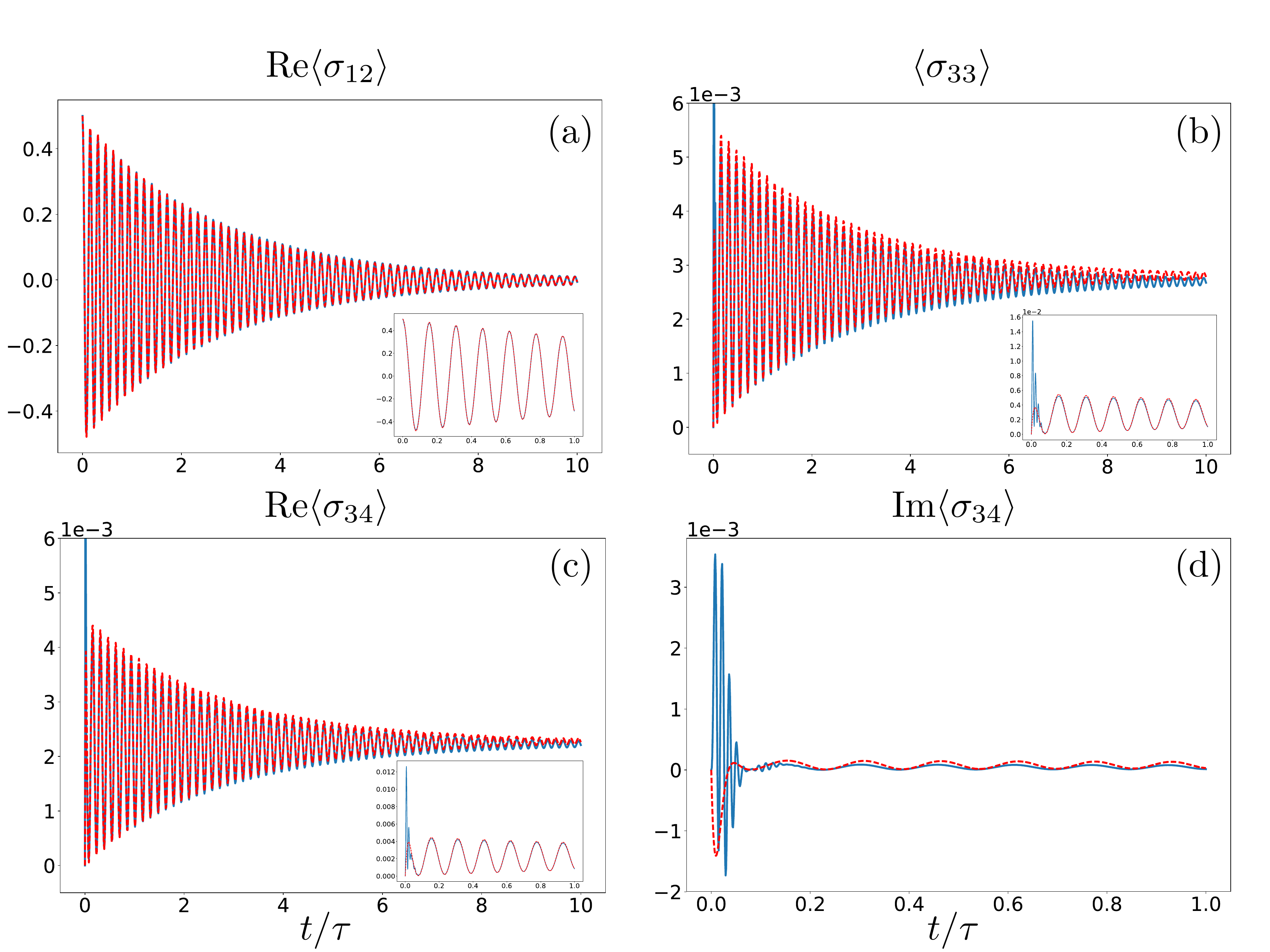}
\caption{Time evolution for the four-level setup with the initial state $|\psi_0\rangle \propto |1\rangle + |2\rangle$ and parameters $\Delta_{13} = 1.0, \Delta_{23} = 0.9, \Delta_{14} = 1.2, \Delta_{24} = 1.1$, $\Omega_{ij} = 0.1$, and $\gamma_{ij} = 0.1$. (a) The real part of $\langle \sigma_{12} \rangle$, (b) $\langle \sigma_{33} \rangle$ inset shows the close-up from $t=0$ to $3\tau$, (c) the real part of $\langle \sigma_{34} \rangle$, and (d) the imaginary part of $\langle \sigma_{34} \rangle$ }
\label{fig:4lv}
\end{figure}

As in the earlier examples, we
find good agreement for the populations and the coherence in the ground state manifold when the 
the state \emph{initially} lies 
there (see Fig.~\ref{fig:4lv}(a) for the real part of the coherence). Unlike in the earlier examples, however, the excited state populations show slight discrepancy as illustrated in Fig.~\ref{fig:4lv}(b) (similar behaviour is observed for $\langle \sigma_{44}\rangle $). The discrepancy at very short times is similar to what we have observed in the 3-level case, but we also see quantitative mismatches that become more pronounced (although still not very large) in the long time limit (the intermediate-time results still match quite well as shown by the inset). We ascribe this to the 
coherence $\langle \sigma_{34} \rangle$ between the excited states [Fig.~\ref{fig:4lv}(c) and (d)], which is created from the states in the lower manifold through 
high-frequency processes. Note, however, that these quantities are orders of magnitudes smaller than the quantities involving ground states.
%-----------------------------Figure 7--------------------------
%-----------------------------Figure 7---------------------------

\begin{figure}[t]
\includegraphics[width=1\columnwidth]{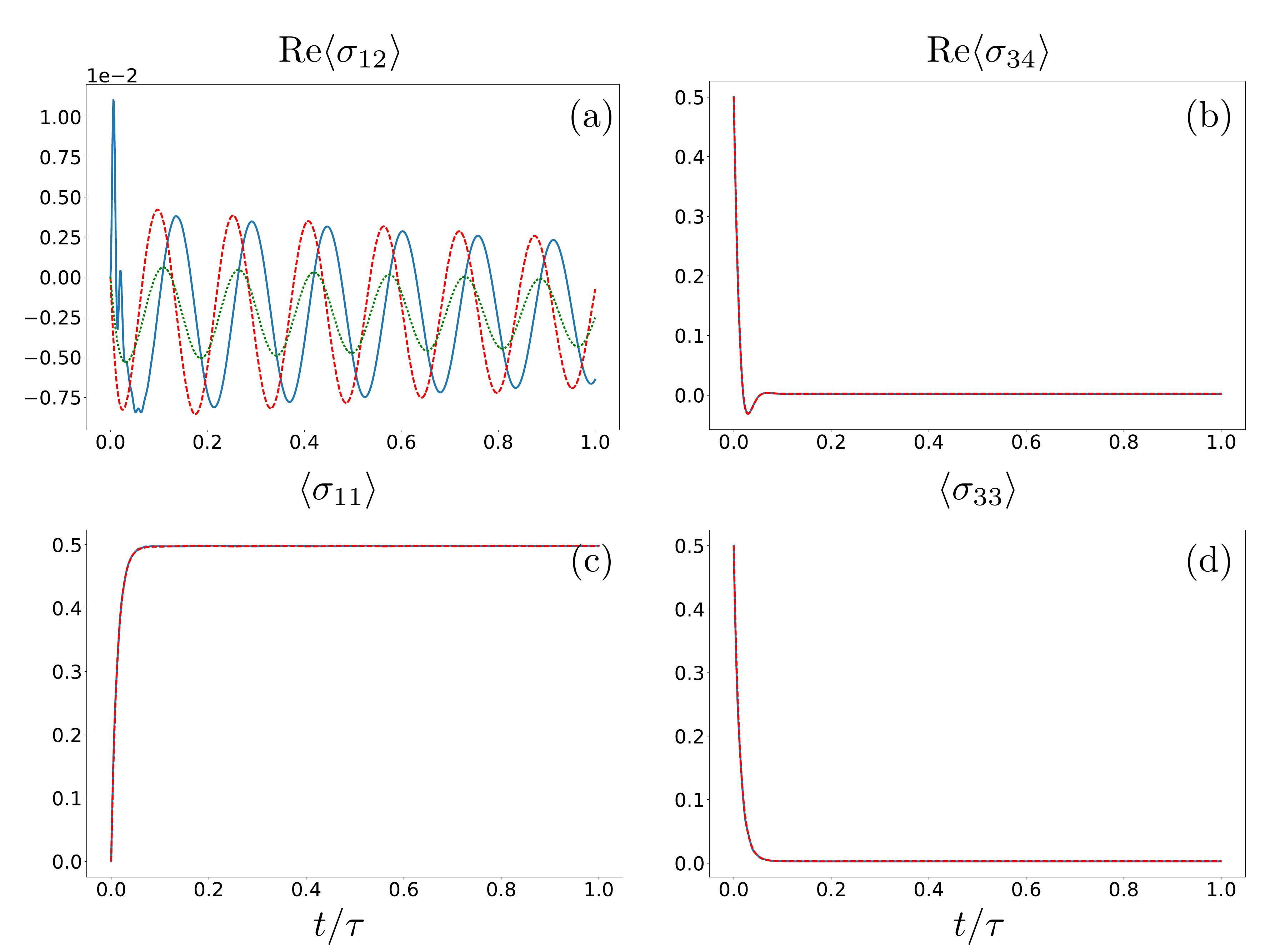}
\caption{Time evolution for the four-level setup with initial state in the excited manifold, $|\psi_0\rangle \propto |3\rangle + |4\rangle$ and parameters $\Delta_{13} = 1.0, \Delta_{23} = 0.9, \Delta_{14} = 1.2, \Delta_{24} = 1.1$, $\Delta_{ij} = 0.1$, and $\gamma_{ij} = 0.1$.  (a) The real part of $\langle \sigma_{12} \rangle$, (b) the real part of $\langle \sigma_{34} \rangle$, (c) $\langle \sigma_{11} \rangle$, and (d) $\langle \sigma_{33} \rangle$.}
\label{fig:4lv_excited}
\end{figure}

Next, we study the other extreme case that the initial state lies entirely in the excited-state manifold.
This time the ground-state coherence is not accurately approximated---although its overall magnitude itself is quite small---whereas the excited-state coherence as well as all the populations are, as shown in Fig.~\ref{fig:4lv_excited}. 
As opposed to all the previous examples, the effective evolutions with (dashed red curves) and without the jump terms (dotted green curves) are somewhat different, albeit only for the ground state coherence; even so the difference is negligible.
The populations show nonoscillatory evolution---indicating that incoherent processes are dominant--- and no transient initial discrepancies 
that are often seen in the earlier examples. 
Lastly, we observe a mixed behaviour for other cases of initial states having comparable occupations in both manifolds. Sometimes, all the low-frequency observables are well-approximated; 
other times, one or more observables, including populations, tend to be poorly approximated.

%-----------------------------Figure 8--------------------------
%-----------------------------Figure 8--------------------------

\section{Conclusion}
\label{sec:conclusion}

We have generalized the time-coarse-grained master equation approach for closed systems 
to open systems and studied its regime of validity by applying it to quantum optical systems of increasing complexity. Specifically, we have considered atomic systems driven by far-detuned lasers, and found that the time-coarse-grained master equation (\ref{eq:meo7}) provides an accurate approximation of the low-energy processes %given by 
produced by the exact open quantum master equation (\ref{eq:meo1}), provided that the following conditions are met: i) the highly oscillating terms are perturbative, i.e., $\Omega \ll \Delta$, in which $\Omega$ is the Rabi frequency due to a driving laser and $\Delta$ is the corresponding laser-atom detuning. We have used $\Omega/\Delta \lesssim 0.1$, but have found that values up to 0.3  give empirically good results; ii) the \emph{initial} state lies mostly either in the lower or in the upper manifold. Increasing the initial coherences between the states in the upper and lower manifolds---a situation at variance with the
underlying assumption of coarse-graining in the first place---leads to low-fidelity approximation of the ground state coherence. 

We have furthermore found that the so-called `jump' decoherence terms can be neglected and the coarse-grained master equation can be reduced to %the simple form
a simpler form \eqref{eq:meo7simple}. 
We notice that this master equation can be obtained from the closed-system version  \eqref{eq:mec3} by replacing $\omega^\pm_{mn}$ with $\tilde{\omega}^\pm_{mn}$ and adding the original Lindblad decoherence terms .
Moreover, our formalism also works when some lasers are resonant (but at least one is far-detuned). In this case, the resonant terms belong to the time-independent Hamiltonian $H_0$ in \eqref{eq:h_eff_o_1}.
In such a case, in turn, a large occupation of a state in the upper manifold is possible, which could then lead to significant `jump' decoherence contributions.

Finally, since our TCG formalism has a hierarchical structure, it is in principle straightforward to expand Eq. \eqref{eq:meo7} to include higher order contributions. 
That is, one can include higher-order terms in the Hamiltonian for stronger driving fields
as well as multiple 
decoherence terms for stronger dissipation effects.

%\section*{acknowledgement}
%C.-W.L. 

\section*{Appendix}
\label{sec:appendix}

%--------------------------------- Appendix  ----------------------------------

%---- Prefix a "A" to all equations, figures, tables and reset the counter ---- 
\setcounter{equation}{0}
%\setcounter{figure}{0}
%\setcounter{table}{0}
%\setcounter{page}{1}
%\makeatletter
\renewcommand{\theequation}{A\arabic{equation}}
%\renewcommand{\thefigure}{A\arabic{figure}}
%\renewcommand{\bibnumfmt}[1]{[A#1]}
%\renewcommand{\citenumfont}[1]{A#1}
%---- Prefix a "A" to all equations, figures, tables and reset the counter ---- 

%\widetext

In this Appendix,
we derive the generic master equation \eqref{eq:meo6} in depth and its detailed form for the four-level system studied in the main text.

\subsection{Detailed derivation of TCG formalism for an open system}
\label{sec:appendixa}

Let us begin from the Dyson expansion of $U$
\begin{eqnarray}
U(t) &=&  \mathds{1} - i \int_{0}^{t} \!\! \, dt_1 H_K(t_1) + (-i)^2 \int_{0}^{t} \!\! dt_1 \!\! \int_{0}^{t_1} \!\! dt_2 \, H_K(t_1) H_K(t_2)  \nonumber \\
& &  + \, \cdots  \nonumber \\
&=&  \mathds{1} + U_1 (t) + U_2 (t) + \, \cdots .
\end{eqnarray}
Its Hermitian conjugate is
\begin{equation}
U^\dag (t) = \mathds{1} + U_1^\dag  (t) + U_2^\dag  (t) + \, \cdots ,
\end{equation}
and, as in the case of closed system \eqref{eq:u_rel_c}, 
\begin{eqnarray}
U^{-1} (t) &=&  \mathds{1} - U_1 (t) + [U_1^2 (t) - U_2 (t)] + \, \cdots , \\
(U^{-1} )^\dag (t) &=&  \mathds{1} - U_1^\dag (t) + [U_1^{\dag 2} (t) - U_2^\dag (t)] + \, \cdots .
\end{eqnarray}
We also get similar relations for the evolution equations of $U_n$'s as \eqref{eq:u_eom_c},
\begin{equation}
i \dot{U}_n(t) = H_K(t) U_{n-1} (t), \quad - i \dot{U}_n^\dag (t) = U_{n-1}^\dag (t) H_K^\dag (t) .
\end{equation}
Using the above relations, we obtain\textcolor{red}{,} for the decay operators $L_{i,U}$ and $\mathcal{J}_\text{int}$, the following:
\begin{eqnarray}
L_{i,U} &=& U^{-1} L_{i,K} U \nonumber\\
&=& [\mathds{1} - U_1 + U_1^2 - U_2 + \cdots] e^{-\gamma_{i,K}t/2} L_i [\mathds{1} + U_1+ U_2 + \cdots] \nonumber\\
&=& e^{-\gamma_{i,K}t/2} L_i + e^{-\gamma_{i,K}t/2} [L_i, U_1] + e^{-\gamma_{i,K}t/2} ([L_i, U_2]  \nonumber\\
&& -U_1 [L_i, U_1]) +  \cdots \nonumber\\
&\equiv& L_{i,0} + L_{i,1} + L_{i,2} +  \cdots,
\end{eqnarray}
and
\begin{eqnarray}
\mathcal{J}_\text{int} &=& \int_{0}^{t} \! \mathcal{J}_\text{tot}  dt' 
\equiv \int_{0}^{t} \! \mathcal{J}_0^U dt'  + \int_{0}^{t} \! \mathcal{J}_1^U dt' + \int_{0}^{t} \! \mathcal{J}_2^U dt' + \cdots \nonumber\\
&\equiv& \mathcal{J}_0 + \mathcal{J}_1 + \mathcal{J}_2 + \cdots ,
\end{eqnarray}
with
\begin{eqnarray}
\mathcal{J}_0^U \rho &\equiv& \sum_i L_{i,0}^{} \, \rho L_{i,0} ^\dag =  \sum_i e^{-\gamma_{i,K}t} L_i\, \rho L_i \\
\mathcal{J}_1^U \rho &\equiv& \sum_i \left(L_{i,0}^{} \, \rho L_{i,1} ^\dag + L_{i,1}^{} \, \rho L_{i,0}^\dag \right)\\
\mathcal{J}_2^U \rho &\equiv& \sum_i \left(L_{i,0}^{} \, \rho L_{i,2} ^\dag + L_{i,1}^{} \, \rho L_{i,1}^\dag + L_{i,2}^{} \, \rho L_{i,0}^\dag \right).
\end{eqnarray}
Now we can identify the detailed forms of $\mathcal{E}_k$ (along with its derivative) and its inverse $\mathcal{F}_k$, 
\begin{align}
% E_0
\mathcal{E}_0 [\rho] &= (1+\overline{\mathcal{J}_0}) \rho = (1+\mathcal{J}_0) \rho , \\
% dE_0/dt
\dot{\mathcal{E}}_0 [\rho] &= \dot{\mathcal{J}}_0 \rho = \mathcal{J}_0^U \rho , \\
% E_1
\mathcal{E}_1 [\rho] &= \overline{\mathcal{J}_1} \rho + \overline{U_1} (1+\mathcal{J}_0) \rho + [(1+\mathcal{J}_0) \rho ] \overline{U_1^\dag} , \\
% dE_1/dt
\dot{\mathcal{E}}_1 [\rho] 
%&= \overline{\mathcal{J}_1^U} \rho - i \overline{H_K} (1+\mathcal{J}_0) \rho + \overline{U_1} \mathcal{J}_0^U \rho + [\mathcal{J}_0^U \rho ] \overline{U_1^\dag} \nonumber\\
%&\quad -i [(1+\mathcal{J}_0) \rho] \overline{H_K^\dag}  ,\\
&= \mathcal{J}_0^U \left(\overline{U_1} \rho + \rho \overline{U_1^\dag} \right) - i\, \overline{H_K} (1+\mathcal{J}_0) \rho \nonumber \\
&\quad +i [(1+\mathcal{J}_0) \rho] \overline{H_K^\dag} \\
% E_2
\mathcal{E}_2 [\rho] &= \overline{\mathcal{J}_2} \rho + \overline{U_2} (1+\mathcal{J}_0) \rho  + \overline{U_1 [ (1+\mathcal{J}_0)\rho] U_1^\dag} \nonumber\\
&\quad  + [(1+\mathcal{J}_0) \rho ] \overline{U_2^\dag}  + \overline{ (\mathcal{J}_1\rho) U_1^\dag} + \overline{U_1 (\mathcal{J}_1} \rho) , \\
% dE_2/dt
\dot{\mathcal{E}}_2 [\rho]  
%&= \overline{\mathcal{J}_2^U} \rho -i\, \overline{H_K U_1} (1+\mathcal{J}_0) \rho + \overline{U_2} \mathcal{J}_0^U \rho + [\mathcal{J}_0^U \rho ] \overline{U_2^\dag} \nonumber\\
%&\quad + i [(1+\mathcal{J}_0) \rho ] \overline{U_1^\dag H_K^\dag} - i \overline{H_K (\mathcal{J}_1} \rho) +\overline{U_1 (\mathcal{J}_1^U} \rho) \nonumber\\
%&\quad + \overline{ (\mathcal{J}_1^U \rho) U_1^\dag} + i\, \overline{ (\mathcal{J}_1\rho) H_K^\dag} -i\,  \overline{H_K [ (1+\mathcal{J}_0)\rho] U_1^\dag} \nonumber\\
%&\quad +  \overline{U_1 [ \mathcal{J}_0^U \rho] U_1^\dag} + i\, \overline{U_1 [ (1+\mathcal{J}_0)\rho] H_K^\dag}, \\
&=  \mathcal{J}_0^U \left(\overline{U_2} \rho + \overline{U_1 \rho U_1^\dag} + \rho \overline{U_2^\dag} \right) -i\, \overline{H_K U_1} (1+\mathcal{J}_0) \rho \nonumber\\
&\; -i\,  \overline{H_K [ (1+\mathcal{J}_0)\rho] U_1^\dag} + i\, \overline{U_1 [ (1+\mathcal{J}_0)\rho] H_K^\dag}, \nonumber\\
&\; + i [(1+\mathcal{J}_0) \rho ] \overline{U_1^\dag H_K^\dag}  - i\, \overline{H_K (\mathcal{J}_1} \rho) + i\, \overline{ (\mathcal{J}_1\rho) H_K^\dag}  \\
% F_0
\mathcal{F}_0 [\rho] &=  (1-\mathcal{J}_0) \rho , \\
% F_1
\mathcal{F}_1 [\rho] &=  -\overline{\mathcal{J}_1} \rho - (1-\mathcal{J}_0) \left(\overline{U_1} \rho + \rho \overline{U_1^\dag} \right) , \\
% F_2
\mathcal{F}_2 [\rho] &=  -\overline{\mathcal{J}_2} \rho - \contraction{}{U}{}{_1 ()\mathcal{J}_1} U_1 (\mathcal{J}_1 \rho)
- \contraction{(}{\mathcal{J}_1}{\rho)}{U_1^\dag} (\mathcal{J}_1 \rho) U_1^\dag
+ \overline{\mathcal{J}_1} \left(\overline{U_1} \rho + \rho \overline{U_1^\dag} \right) \nonumber \\
&\quad + (1-\mathcal{J}_0)\, \Bigg[ \left(\overline{U_1}^2 - \overline{U_2}\right) \rho + \rho \left(\overline{U_1^\dag}^2 - \overline{U_2^\dag }\right) \nonumber\\
&\quad + \overline{U_1^{}} \rho \overline{U_1^\dag} - \contraction{}{U}{_1\rho}{U} U_1 \rho U_1^\dag \Bigg].
\end{align}
From these formulas, we obtain the time-coarse-grained master equation,
\begin{eqnarray}
i\, \dot{\overline{\rho}}_K (t) &=& i \dot{\mathcal{E}}[\rho_0] = i \dot{\mathcal{E}}[\mathcal{F}[\overline{\rho}_K (t)]] 
= i \sum_{k=0}^{\infty} \sum_{j=0}^{k} \dot{\mathcal{E}}_j [\mathcal{F}_{k-j}[\overline{\rho}_K (t)]]  \nonumber\\
&\equiv& \sum_{k=0}^{\infty} \mathcal{L}_k [\overline{\rho}_K (t)].
\label{eq:meo5}
\end{eqnarray}
where, up to the 2nd order, the superoperators $\mathcal{L}_k$ are given as
\begin{eqnarray}
\mathcal{L}_0 [\rho] &=& i \mathcal{J}_0^U \rho , \\
\mathcal{L}_1 [\rho] &=& \overline{H_K} \rho - \rho \overline{H_K^\dag} ,\\
\mathcal{L}_2 [\rho] &=& \contraction{}{H}{_K}{U} H_K U_1 \rho+ \contraction{}{H}{_K \rho}{U} H_K \rho U_1^\dag - \contraction{}{U}{_1 \rho}{H} U_1 \rho H_K^\dag - \contraction{\rho}{U}{_1}{H} \rho U_1^\dag H_K^\dag \nonumber\\
&& + \contraction{}{H}{_K(}{\mathcal{J}} H_K (\mathcal{J}_1 \rho) - \contraction{(}{\mathcal{J}}{_1 \rho)}{H} (\mathcal{J}_1 \rho) H_K^\dag .
\end{eqnarray}
In order to return to the original frame, we apply the inverse of the transform \eqref{eq:transform_k} to \eqref{eq:meo5} (also up to the 2nd order), which reduces to
\begin{eqnarray}
i \dot{\overline{\rho}}(t) 
&=& 
-i \sum_{i} \mathcal{K}_{L_i} \overline{\rho} +
e^{-K t} \Big(\mathcal{L}_0[\overline{\rho}_K] + \mathcal{L}_1[\overline{\rho}_K] + \mathcal{L}_2[\overline{\rho}_K] \Big) e^{-K t} \nonumber\\
&=& i \sum_{i} \left( \mathcal{J}\!_{L_i} \rho - \mathcal{K}_{L_i}  \rho \right) + 
[\overline{H},\overline{\rho}] + 
\contraction{}{H}{}{U} H \tilde{U}_{1} \overline{\rho} + 
\contraction{}{H}{\overline{\rho}}{U} H \overline{\rho} \tilde{U}_1^\dag \nonumber\\
&& - \contraction{}{U}{_1 \rho}{H} \tilde{U}_1 \overline{\rho} H
- \contraction{\overline{\rho}}{U}{_1}{H} \overline{\rho} \tilde{U}_1^\dag H 
+ \contraction{}{H}{(}{\mathcal{J}} H (\tilde{\mathcal{J}}_1 \overline{\rho}) 
- \contraction{(}{\mathcal{J}}{_1 \overline{\rho})}{H} (\tilde{\mathcal{J}}_1 \overline{\rho}) H . \nonumber
\label{eq:meo8}
\end{eqnarray}
Rearranging to collect all the Hamiltonian-like terms gives the final master equation in the main text.

\subsection{Detailed derivation of the coarse-grained master equation for the four-level System}
\label{sec:appendixb}

In this section, we derive the TCG master equation for the four-level system given in the main text. 
To start with, note that
$\overline{H} = 0$, $\overline{\tilde{U}_1} = -(h_{ij} - h_{ij}^\dagger)/\tilde{\Delta}_{ij}$, and
\begin{align}
\overline{H\tilde{U}_1} = \sum_{\langle i,j \rangle}\sum_{\langle k,l\rangle} \frac{h_{ij} h_{kl}^\dagger}{\tilde{\Delta}_{kl}} e^{i(\Delta_{ij} - \Delta_{kl})t} - \frac{h_{ij}^\dagger h_{kl} }{\tilde{\Delta}_{kl}}e^{-i(\Delta_{ij}-\Delta_{kl})t},
\end{align}
and hence
\begin{align}
H_{\rm eff} &= \frac{1}{2}\left(\overline{H\tilde{U}_1} + \overline{\tilde{U}_1^\dagger H}  \right) \nonumber \\
&= \sum_{\langle i,j \rangle}\sum_{\langle k,l\rangle}\frac{1}{2}\left( \frac{1}{\tilde{\Delta}_{kl}} + \frac{1}{\tilde{\Delta}_{ij}^*}\right)e^{i(\Delta_{ij} - \Delta_{kl})t}  \left[ h_{ij} h_{kl}^\dagger  - h_{ij}^\dagger h_{kl} \right].
\end{align}
If we let $\langle i, j\rangle = m$, $\langle k,j\rangle = n$, and $h_{ij} = h_{ij}e^{i\Delta_{ij}t}$, the equation reduces to
\begin{align}
H_{\rm eff} &= \sum_{m,n} \frac{1}{2}\left(\frac{1}{\tilde{\Delta}_{n}} +\frac{1}{\tilde{\Delta}_{m}^*}\right)[h_m(t)h_n^\dag(t) - h_m^\dagger(t) h_n(t)] \nonumber \\ &\equiv \sum_{mn} \frac{1}{\tilde{\Delta}_{nm}^+}[h_m(t)h_n^\dag(t) - h_m^\dagger(t) h_n(t)] 
\end{align}
in obvious agreement with Eq.~(\ref{eq:h_eff_o_2}). 
Next,
\begin{align}
H_{\rm eff} &= \frac{1}{2}\left(\overline{H\tilde{U}_1} - \overline{\tilde{U}_1^\dagger H}  \right) \nonumber \\
&= \sum_{\langle i,j \rangle}\sum_{\langle k,l\rangle}\frac{1}{2}\left( \frac{1}{\tilde{\Delta}_{kl}} - \frac{1}{\tilde{\Delta}_{ij}^*}\right)e^{i(\Delta_{ij} - \Delta_{kl})t}  \left[ h_{ij} h_{kl}^\dagger  - h_{ij}^\dagger h_{kl} \right]
\end{align}
and
\begin{align}
\overline{H\overline{\rho}\tilde{U}^\dagger_1} - \overline{\tilde{U}_1\overline{\rho}H} &  \nonumber \\
=\sum_{\langle i,j \rangle}\sum_{\langle k,l\rangle}  & \left[  \frac{1}{\tilde{\Delta}_{kl}^*} \left( h_{ij}^\dagger (t) \overline{\rho} h_{kl}(t)  - h_{ij}(t) \overline{\rho} h_{kl}^\dagger(t)  \right) \right. \nonumber \\
 & - \left. \frac{1}{\tilde{\Delta}_{ij}} \left( h_{ij}^\dagger (t) \overline{\rho} h_{kl}(t)  - h_{ij}(t) \overline{\rho} h_{kl}^\dagger(t)  \right) \right] \nonumber \\
 =\sum_{\langle i,j \rangle}\sum_{\langle k,l\rangle} & \left[\left( \frac{1}{\tilde{\Delta}_{ij}} - \frac{1}{\tilde{\Delta}_{kl}^*}\right) h_{ij}(t) \overline{\rho} h_{kl}^\dag(t)\right.  \nonumber \\
 &- \left. \left( \frac{1}{\tilde{\Delta}_{ij}} - \frac{1}{\tilde{\Delta}_{kl}^*}\right) h_{ij}^\dag (t) \overline{\rho} h_{kl}(t) \right].
\end{align}
Making the replacements of $\langle i,j\rangle$ and $\langle k,l\rangle$ to $m$ and $n$ again, we obtain
\begin{align}
\frac{1}{2}&\left( \overline{H\tilde{U}_1} - \overline{\tilde{U}_1^\dagger H}\right)\overline{\rho} + \overline{\rho}\frac{1}{2}\left( \overline{H\tilde{U}_1} - \overline{\tilde{U}_1^\dagger H}\right)  + \overline{H\overline{\rho}\tilde{U}^\dagger_1} - \overline{\tilde{U}_1\overline{\rho}H}   \nonumber \\
&= \sum_{m,n}\frac{2}{\tilde{\Delta}_{mn}^-} \left( \mathcal{D}_{h_m(t),h_n^\dag(t)} \overline{\rho}  - \mathcal{D}_{h_m^\dag(t),h_n(t)} \overline{\rho} \right),
\end{align}
in agreement with the second line of Eq.~(\ref{eq:meo7}).

The last ingredient that needs to be computed is
\begin{align}
\mathcal{J}_1\rho = \int_0^t dt' \mathcal{J}_1^U\rho = \int_0^t dt' \sum_i L_{i,0}\rho L_{i,1}^\dag + L_{i,1}\rho L_{i,0}^\dag.
\end{align}
Here, $L_{i,0} = e^{-\gamma_{i,K}t'/2}L_i$ and $L_{i,1} = e^{-\gamma_{i,K}t'/2}[L_i,U_1]$, and hence
\begin{align}
\mathcal{J}_1\rho = &\int_0^t dt' \sum_n e^{-\gamma_{n,K}t'} \left( L_n\rho[L_n,U_1]^\dag + [L_n,U_1]\rho L_n^\dag \right) \nonumber \\
= &\int_0^t dt' \sum_n e^{-\gamma_{n,K}t'} L_n \rho \sum_{\langle j,k \rangle}\frac{1}{\tilde{\Delta}_{jk}^*}\left[ e^{-i\tilde{\Delta}_{jk}^*t'}h_{jk}^\dag - e^{i\tilde{\Delta}_{jk}^*t'}h_{jk} \right. \nonumber \\ & \left. + h_{jk}^\dag - h_{jk},L_n^\dag \right]  + {\rm H.c.} \nonumber \\
=& \sum_{n, \langle j,k \rangle}  \frac{i}{\tilde{\Delta}_{jk}^* (\tilde{\Delta}_{jk}^* -i \gamma_{n,K})} e^{-i(\tilde{\Delta}_{jk}^*-i\gamma_{n,K})t} L_n\rho\left[ h_{jk}^\dagger, L_n^\dagger\right] \nonumber \\
&\,\,\,\,\,-   \frac{i}{\tilde{\Delta}_{jk}^* (\tilde{\Delta}_{jk}^* +i \gamma_{n,K})} e^{i(\tilde{\Delta}_{jk}^*+i\gamma_{n,K})t}L_n\rho\left[ h_{jk}, L_n^\dagger\right] \nonumber \\
&\,\,\,\,\,+ {\rm terms \,not \, containing \,} e^{\pm i \tilde{\Delta}_{jk}^*t} \nonumber \\
&+ {\rm H.c.}.
\end{align}
The terms that has the time dependence $e^{-\gamma_{n,K}t}$ and time-independent terms will be coarse-grained out later when we compute $\overline{H\tilde{\mathcal{J}}_1\bar{\rho}}$ and $\overline{\left( \tilde{\mathcal{J}}_1\bar{\rho}\right)H}$. Noting that $\left[ h_{jk}, L_n \right] = 0$ for all $n, j$, and $k$ and upon returning to the non-decaying frame, one obtains
\begin{align}
\tilde{\mathcal{J}}_1\rho &= \sum_{n,\langle j,k\rangle} -   \frac{i}{\tilde{\Delta}_{jk}^* (\tilde{\Delta}_{jk}^* +i \gamma_{n,K})}L_n\rho\left[ h_{jk}(t), L_n^\dagger\right] + {\rm H.c.} \nonumber \\
&+ {\rm irrelevant\,terms.}
\end{align}
From these, the jump-induced terms are  readily calculated as
\begin{align}
i \overline{H  \tilde{\mathcal{J}}_1\bar{\rho}}   =&\sum_{n, \langle j,k \rangle, \langle l,m \rangle} \frac{1}{\tilde{\Delta}_{jk}^*(\tilde{\Delta}_{jk}^* +i \gamma_{n,K})} h_{lm}^\dagger (t) L_n\bar{\rho} \left[ h_{jk}(t), L_n^\dagger \right]  \nonumber \\
&  +\frac{1}{\tilde{\Delta}_{jk}(\tilde{\Delta}_{jk} -i \gamma_{n,K})} h_{lm}(t) \left[ h_{jk}(t), L_n^\dagger \right]^\dagger \bar{\rho} L_n^\dagger ,
\end{align}
and
\begin{align}
i \overline{ \tilde{\mathcal{J}}_1 \bar{\rho} H} =&\sum_{n, \langle j,k \rangle, \langle l,m \rangle} \frac{1}{\tilde{\Delta}_{jk}^*(\tilde{\Delta}_{jk}^* +i \gamma_{n,K})}  L_n\bar{\rho} \left[ h_{jk}(t), L_n^\dagger \right] h_{lm}^\dagger (t)  \nonumber \\
& +\frac{1}{\tilde{\Delta}_{jk}(\tilde{\Delta}_{jk} -i \gamma_{n,K})} \left[ h_{jk}(t), L_n^\dagger \right]^\dagger \bar{\rho} L_n^\dagger h_{lm}(t) .
\end{align}
From which we obtain the total `jump' contributions to the first order:
\begin{align}
i &\Big[ \overline{ \tilde{\mathcal{J}}_1 \bar{\rho} H} - \overline{H  \tilde{\mathcal{J}}_1\bar{\rho}} \Big] = \nonumber \\
&\sum_{n, \langle j,k \rangle, \langle l,m \rangle}  \frac{1}{\tilde{\Delta}_{jk}^*(\tilde{\Delta}_{jk}^* +i \gamma_{n,K})} \left[ L_n\bar{\rho} \left[ h_{jk}(t), L_n^\dagger \right],  h_{lm}^\dagger (t) \right] + {\rm H.c.}.
\end{align}
Using shorthand notations $h(t) = \sum_{\langle j,k \rangle} h_{jk}(t)$ and
$h_{\Delta,n}(t) = \sum_{\langle j,k \rangle} h_{jk}(t)/(\tilde{\Delta}_{jk}^*(\tilde{\Delta}_{jk}^*+i\gamma_{n,K}))$,
we can condense the above expression as
\begin{align}
i \Big[ \overline{ \tilde{\mathcal{J}}_1 \bar{\rho} H} - \overline{H  \tilde{\mathcal{J}}_1\bar{\rho}} \Big] = \sum_{n} \left[ L_n\bar{\rho} \left[ h_{\Delta,n}(t), L_n^\dagger \right],  h^\dagger (t) \right] + {\rm H.c.}.
\end{align}

\end{document}